\documentclass[12pt]{article}
\usepackage{UF_FRED_paper_style}
\usepackage{times}
\usepackage{amsmath, amssymb}
\usepackage{geometry}
\usepackage{cancel}
\usepackage{marginnote}
\usepackage{amsthm}
\usepackage{cleveref}
\usepackage{natbib}
\usepackage{appendix}
\usepackage{booktabs}

\hypersetup{
    colorlinks = true,
    citecolor = blue
}
\usepackage{lipsum}  
\usepackage{setspace}

\usepackage{bm}
\usepackage{booktabs}
\usepackage{subcaption}
\usepackage{graphicx}
\title{\textbf{Bayesian Power and Sample Size Calculations for Bayes Factors in the Binomial Setting}}

\author{Riko Kelter \thanks{Correspondence concerning this article should be addressed to riko.kelter@uni-siegen.de
    Draft version 1.0, 29/01/25. This paper has not been peer reviewed. Please do not copy or cite without author's permission. Data and R code to reproduce our results are openly available at \url{https://osf.io/mn4e3/}. We declare no conflict of interest.}\\
	Department of Mathematics\\
	University of Siegen\\
    \and Samuel Pawel \\
    Epidemiology, Biostatistics and Prevention Institute\\
    Center for Reproducible Science \\
    University of Zurich}

\date{\today}

\begin{document}

{\setstretch{.8}
\maketitle
\begin{abstract}
Bayesian design of experiments and sample size calculations usually rely on complex Monte Carlo simulations in practice. Obtaining bounds on Bayesian notions of the false-positive rate and power therefore often lack closed-form or approximate numerical solutions. In this paper, we focus on the sample size calculation in the binomial setting via Bayes factors, the predictive updating factor from prior to posterior odds. We discuss the drawbacks of sample size calculations via Monte Carlo simulations and propose a numerical root-finding approach which allows to determine the necessary sample size to obtain prespecified bounds of Bayesian power and type-I-error rate almost instantaneously. Real-world examples and applications in clinical trials illustrate the advantage of the proposed method. We focus on point-null versus composite and directional hypothesis tests, derive the corresponding Bayes factors, and discuss relevant aspects to consider when pursuing Bayesian design of experiments with the introduced approach. In summary, our approach allows for a Bayes-frequentist compromise by providing a Bayesian analogue to a frequentist power analysis for the Bayes factor in binomial settings. A case study from a Phase II trial illustrates the utility of our approach. The methods are implemented in our R package \texttt{bfpwr}.


\noindent

\end{abstract}
}

\section{Introduction}\label{sec:intro}
To ensure efficient and ethical use of resources, sample size calculations are an essential component of experimental design in both frequentist and Bayesian statistics.
Accurate planning of the sample size is particularly important in the context of hypothesis testing. There, a prespecified power under the alternative hypothesis $H_1$ is desired and the probability of incorrectly rejecting the null hypothesis $H_0$ needs to be bounded, which often resembles the conventional $\alpha$-level of $0.05$ \citep{Bijma2017}, or $0.025$ when a one-sided test is used \citep{Matthews2006}. While there exist a plethora of formulas for sample size calculation for standard frequentist tests such as the $t$ test or chi-squared test \citep{Matthews2006}, the situation is not this simple for Bayesian statisticians.

Bayesian statistics is first and foremost troubled by the existence of various indices for hypothesis testing \citep{Kelter2021}. For example, one might decide to use posterior probabilities $P(H_0 \mid y)$ and $P(H_1 \mid y)$ given the data to test $H_0$ versus $H_1$. Alternatively, one could opt for the Bayes factor $\mathrm{BF}_{01}(y):=\frac{f(y \mid H_0)}{f(y \mid H_1)}$, the predictive updating factor from prior to posterior odds:
\begin{align}\label{eq:bayesFactor}
    \underbrace{\frac{P(H_0 \mid y)}{P(H_1 \mid y)}}_{\text{Posterior odds}} = 
    \underbrace{\frac{f(y \mid H_0)}{f(y \mid H_1)}}_{\text{Bayes factor}} \cdot \underbrace{\frac{P(H_0)}{P(H_1)}}_{\text{Prior odds}}
\end{align}
Sample size planning and the resulting power under $H_1$ can (and often does) vary for different approaches to testing, such as posterior probabilities and Bayes factors \citep{Kelter2020}.

However, the more pressing problem with Bayesian sample size calculations is that in contrast to frequentist hypothesis testing,  there exist little to no closed-form or at least numerical solutions. The question how large a sample size should be to provide a power of e.g. $80\%$, and a type-I-error rate of e.g. $5\%$, must therefore often rely on simulation (compare \Cref{sec:MCse}). Although power and type-I-error rates are no original Bayesian concepts per se, the idea to compute frequentist metrics of Bayesian indices for hypothesis testing is widely present in the literature \citep{Pourmohammad2023,Grieve2022,Rosner2021,Kleijn2022,Weiss1997,DeSantis2004} and ranges back to \cite{Good1983a,Good1960} and \cite{Kerridge1963}. Further perspectives on such Bayes-frequentist compromises in the sense of ``frequentistically calibrated'' Bayesian approaches are given by \cite{Dawid1982}, \cite{Rubin1984}, \cite{Little2006} and \cite{Grieve2016}. The calibrated Bayes approach is often also recommended by health authorities, for example, the U.S. Food and Drug Administration accepts Bayesian analysis if they are appropriately calibrated in a frequentist sense \citep{FDABayes2010}.

Such a hybrid perspective on Bayesian design of experiments takes the stance to investigate the long-term behaviour of Bayesian quantities like posterior probabilities or Bayes factors, to compute Bayesian versions of power and type-I-error such as
\begin{align}\label{eq:bayesianPower}
    P(\mathrm{BF}_{01}(y) < k \mid H_0) \hspace{1cm} \text{and} \hspace{1cm} P(\mathrm{BF}_{01}(y) < k \mid H_1)
\end{align}
For example, for $k=1/10$ -- the usual threshold for strong evidence according to the scale of \cite{Jeffreys1939} -- when $P(\mathrm{BF}_{01}(y) < k \mid H_0)\leq \alpha$ holds, the probability to obtain a Bayes factor indicating strong evidence against the null hypothesis (that is, in favour of the alternative $H_1$) is bounded by $\alpha$ under $H_0$.\footnote{This holds because $P(\mathrm{BF}_{01}(y) < k \mid H_0)=P(1/\mathrm{BF}_{01}(y) > 1/k \mid H_0)=P(\mathrm{BF}_{10}(y) >10 \mid H_0) \leq \alpha$ and $\mathrm{BF}_{10}(y) > 10$ amounts to the Bayes factor in favour of $H_1$ indicating (at least) strong evidence for $H_1$. Thus, $P(\mathrm{BF}_{01}(y) < k \mid H_0)$ resembles an intuitive Bayesian type-I-error rate.} Thus, for $\alpha:=0.05$ the Bayesian type-I-error rate is controlled at $5\%$ then. Likewise, if $P(\mathrm{BF}_{01}(y) < k \mid H_1)>1-\beta$ for, say $\beta:=0.2$ and $k=1/10$, the probability to obtain strong evidence in favour of $H_1$, when $H_1$ is true, is $80\%$.\footnote{This holds because $P(\mathrm{BF}_{01}(y) < k \mid H_1)=P(\mathrm{BF}_{10}(y) > 1/k \mid H_1)=P(\mathrm{BF}_{10}(y) > 10 \mid H_1)$.} A critical goal in Bayesian design of experiments and sample size calculation therefore is to provide solutions to the inequalities
\begin{align}\label{eq:bayesianDesign}
    P(\mathrm{BF}_{01}(y) < k \mid H_0)\leq \alpha \hspace{1cm} \text{and} \hspace{1cm} P(\mathrm{BF}_{01}(y) < k \mid H_1)> 1-\beta
\end{align}
for some $\alpha$ and $\beta$ in $(0,1)$.  We close the introduction with two comments. \begin{itemize}
    \item[$\blacktriangleright$]{We decided to use thresholds $k<1$ in the conditions \Cref{eq:bayesianDesign} because this is the original Bayes factor orientation of \cite{Jeffreys1939} who also used $\mathrm{BF}_{01}$ instead of $\mathrm{BF}_{10}$. Therefore, we opt for using the Bayes factor in favour of $H_0$ in all our conditions and formulations, in particular, in the notation of a Bayesian type-I-error rate and Bayesian power as shown in \Cref{eq:bayesianDesign}. However, if desired, all equations and results can be easily converted to the $\mathrm{BF}_{10}$ orientation by taking the reciprocal of the Bayes factor and $k$ and inverting inequalities.}
    \item[$\blacktriangleright$]{A further advantage of this notation is that small values of $\mathrm{BF}_{01}$ indicate evidence against the null hypothesis $H_0$, which is in line with frequentist reasoning when using \textit{p}-values. Accommodating our approach should therefore cause no trouble for frequentists who want to use a Bayesian design with desirable frequentist properties. Note that the latter property does not hold for $\mathrm{BF}_{10}$, where small values indicate evidence \textit{in favour} of the null hypothesis $H_0$.}
\end{itemize} 

\subsection{The curse of the Monte Carlo standard error}
\label{sec:MCse}
As noted above, there are little to no closed-form formulas or at least numerical solutions in almost all practically relevant settings, which allow to compute \Cref{eq:bayesianPower} in a simple manner with some notable exceptions such as $z$ and $t$ test Bayes factors \citep{Weiss1997,DeSantis2004, PawelHeld2024,WongTendeiro2024}. Thus, the status quo in Bayesian design of experiments and sample size planning often relies on simulation, and the usual standard consists of conducting a Monte Carlo simulation to investigate the quantities in \Cref{eq:bayesianPower} \citep{Schoenbrodt2017,Schonbrodt2018,Stefan2022a,Grieve2022,Kelter2020,Kelter2021c,Kelter2021d,Kelter2021BMCHodgesLehmann,Makowski2019,KelterSchnurr2024,Berry2006,Berry2011,GelfandWang2002}. In this section, we outline a principal problem connected with this approach.


Monte Carlo simulations are abundant in Bayesian sample size calculation in the context of hypothesis testing \citep{Berry2011,Grieve2022, Schoenbrodt2017}. They typically proceed as follows:
First, simulation of a parameter value $\theta_i$, $i=1,...,n_{sim}$ under $H_0$ or $H_1$ is necessary. Then, for each parameter value $\theta_i$, a data set $Y^i:=y_{1}^i,....,y_{n}^i$ of size $n$ needs to be simulated, so there are $n_{sim}$ data sets of sample size $n$. 
Subsequently, the computation of the quantity of interest (e.g. the Bayes factor) based on these $n_{sim}$ realizations $Y^i$ follows. Due to the strong law of large numbers, the Monte Carlo average then converges for $n_{sim}\rightarrow \infty$ against the desired probability \citep{Robert2004}, for example:
\begin{align}\label{eq:monteCarloEstimate}
    \frac{1}{n_{sim}}\sum_{i=1}^{n_{sim}} 1_{\{\mathrm{BF}_{01}(Y^i)<  k \}} \xrightarrow[n_{sim}\rightarrow \infty]{P_{H_0}} P(\mathrm{BF}_{01}<  k \mid H_0)
\end{align}
where we suppose that the parameter draws $\theta_i$ are simulated according to $H_0$ in \Cref{eq:monteCarloEstimate}. Here, $1_{\{ \mathrm{BF}_{01}(Y^i)<  k \}}$ denotes the indicator function, which takes value $1$ if $\mathrm{BF}_{01}(Y^i)< k$ and else $0$. This Monte Carlo estimate provides an estimate for the fixed sample size $n$ based on $n_{sim}$ simulated data sets. In practice, the threshold $k$ and sample size $n$ need to be balanced to fulfill \Cref{eq:bayesianDesign}. However, a reliable bound on the Bayesian type-I-error (and likewise, Bayesian power) is obtained only, if $n_{sim}$ is large enough.

A problem now occurs as Monte Carlo uncertainty enters the stage. As noted in \cite{Morris2019}, Monte Carlo standard errors are essential to quantify the uncertainty of a Monte Carlo estimate due to the finiteness of the simulation size $n_{sim}$. For example, using $n_{sim}=1'000$ simulated realizations of the data of sample size $n=100$ under $H_0$ will produce a less reliable estimate of $P(\mathrm{BF}_{01}(y)< k \mid H_0)$ than using $n_{sim}=10'000$ simulated realizations. Providing a Monte Carlo standard error is therefore quintessential to indicate the uncertainty due to simulation, a proposal that was also included in the \textit{Bayesian Simulation Study} (BASIS) framework of \cite{Kelter2023}. 
However, guidelines and tutorials in Bayesian Monte Carlo simulation \citep[e.g.][]{Kruschke2021, Depaoli2017, VandeSchoot2021, Schoenbrodt2017, Stefan2019} 
often recommend fixed values such as $n_{sim}=10'000$. This is clearly inadequate, as in one case $n_{sim}=10'000$ may be enough, while in another it may result in an unacceptably large Monte Carlo standard error. Even when $n_{sim}=10'000$ may be enough, it could be ``too much'' in the sense that $n_{sim}=5'000$ would produce a sufficiently small Monte Carlo standard error and the computational resources could have been used more effectively otherwise.

Taking stock, Monte Carlo simulations provide a working but computationally expensive tool for Bayesian design of experiments and sample size calculations. The Monte Carlo uncertainty inherent to these approaches troubles Bayesian sample size calculations as it often undermines the reliability of a simulation-based result when not taken carefully into account.\footnote{We emphasize that Monte Carlo simulations are of course helpful and necessary when no other solutions are available, but simultaneously stress that Monte Carlo standard errors must then accompany the Monte Carlo estimate. We omit details on the computational approaches to compute a Monte Carlo standard error (which include Bootstrap resampling, see the appendix in \cite{Kelter2023}) and only note that they usually add another layer of complexity and computational effort to the possibly already demanding simulation study. } 
Currently, there is a lack of other approaches such as closed-form solutions or numerical approximations to solve \Cref{eq:bayesianDesign}. We now turn to our solution to this problem and provide an outline of what this paper contributes to the existing literature.

\subsection{Outline}
\begin{figure}[!htb]
    \centering
    \includegraphics[width=1.0\linewidth]{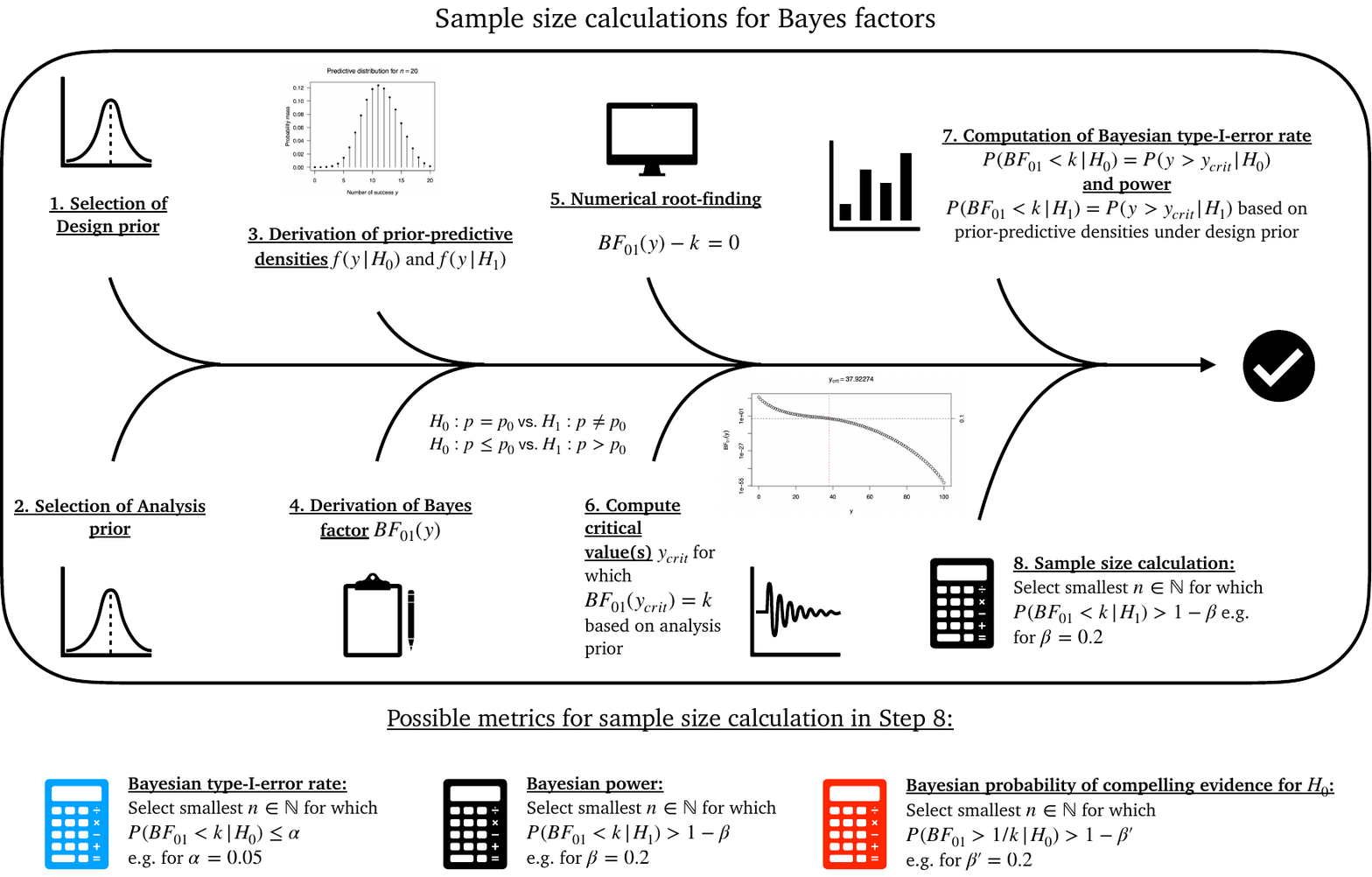}
    \caption{Overview of the root-finding based approach to Bayesian sample size calculations for Bayes factors.}
    \label{fig:overview}
\end{figure}

In this paper, we propose a sample size calculation procedure for Bayes factors in the binomial setting outlined in the next section and already discussed briefly in \Cref{sec:intro}. Figure \ref{fig:overview} provides an overview about the approach, and while the idea is generalizable to other settings such as $t$ tests, $z$ tests \citep{PawelHeld2024, WongTendeiro2024} and other situations, we restrict the focus of the current manuscript to the binomial setting. The latter is important in phase II clinical trials with binary endpoint, where the proof of concept of a novel drug is studied. Other applications are given in \Cref{sec:examples}.

The plan of the paper is as follows: \Cref{sec:rootfinding} outlines the main idea and key steps depicted in \Cref{fig:overview}. We detail the two most relevant hypothesis tests in the binomial model, that is, the point-null versus composite test $H_0:p=p_0$ versus $H_1:p\neq p_0$ and the directional test of $H_0:p\leq p_0$ versus $H_1:p > p_0$. We obtain the one-sided test of $H_0:p=p_0$ versus $H_1:p > p_0$ as a special case under our chosen priors. Therefore, we detail the derivation of the Bayes factor and walk through the entire process depicted in \Cref{fig:overview}. We illustrate the idea in Section \ref{sec:examples} and provide real world examples. The ultimate goal as shown in \Cref{fig:overview} is to obtain a calibrated Bayesian experimental design in the sense that one or both inequalities in \Cref{eq:bayesianDesign} are fulfilled. 
In the following section \Cref{sec:powerConcepts} we also discuss alternatives to this frequentist approach in the sense that other metrics could be of interest to a Bayesian, such as the probability to confirm $H_0$ when it is indeed true.
We close in \Cref{sec:discussion} with a discussion and outlook for future research.

\section{The root-finding approach to Bayesian sample size calculation}\label{sec:rootfinding}

\subsection{One-sided hypothesis test}

Our main assumption here is that the observed data random variable $Y$ follows a binomial distribution with parameters $n$ and $p$, $Y\sim \mathrm{Bin}(n,p)$. Its probability mass function is given as
\begin{align}\label{eq:binomialLik}
    f(y \mid p)={n\choose y}p^y (1-p)^{n-y}
\end{align}
for $y=0,...,n$. In this setting, we have $p\in [0,1]=:\Theta$, so the parameter space $\Theta$ is the unit interval. First, a prior $P_p$ is chosen for the parameter $p \in [0,1]$. The $\mathrm{Beta}(a_0,b_0)$ distribution is a conjugate prior for the binomial likelihood, and when chosen as the prior, the posterior $P_{p \mid Y}$ is also Beta-distributed \citep{Held2014}:
\begin{align*}
	p \mid Y=y\sim \mathrm{Beta}(a_0+y,b_0+n-y)
\end{align*}
As a consequence, a natural choice for a prior $P_p$ is the beta distribution. We first deal with the one-sided hypothesis test of
\begin{align}\label{eq:binomialOneSidedTest}
    H_0:p\leq p_0 \hspace{1cm} \text{ versus } \hspace{1cm} H_1:p>p_0
\end{align}
for some $p_0\in [0,1]$. This situation arises for example in single-arm clinical phase II studies which aim at demonstrating the efficacy of a potential novel drug or therapy \citep{Zhou2017,KelterSchnurr2024}.

\subsubsection{Choice of the design and analysis prior}

The first step for the sample size calculation approach is shown as steps 1 and 2 in \Cref{fig:overview}. Thus, a so called design prior and analysis prior must be chosen \citep{OHagan2001}. We denote the former by $P^d$ and the latter by $P^{a}$. The idea of the design prior is that we base the planning of the sample size and properties such as the Bayesian type-I-error rate and power -- compare \Cref{eq:bayesianPower} -- on this prior $P^d$. However, the planning stage might include subjective beliefs, so the analysis -- in this case a Bayes factor test -- should be based on another (more objective) prior distribution $P^{a}$.\footnote{These objective beliefs could result in a design that does not meet our demands formulated in \Cref{eq:bayesianDesign}.} When choosing $P^d = P^{a}$, the design and analysis prior agree and we base both the planning and analysis of the study on the same distribution.

Therefore, we start with the design prior and must choose a prior under $H_0$ and under $H_1$, to use the prior-predictive density in a second step to calculate our desired quantities in \Cref{eq:bayesianDesign}. Therefore, we choose truncated Beta priors both under $H_0$ and $H_1$ as follows:
\begin{align}\label{eq:designPriorH0OneSidedTesting}
    &p \mid H_0 \sim \mathrm{Beta}(a_d,b_d)_{[0,p_0]}
\end{align}
where $\mathrm{Beta}(a_d,b_d)_{[0,p_0]}$ denotes the truncated Beta distribution to the interval $[0,p_0]$ with density
\begin{align}\label{eq:truncatedBetaDensity}
    f(p \mid a_d,b_d)=\frac{p^{a_d-1}(1-p)^{b_d-1}}{\mathrm{B}(a_d,b_d)(I_{p_0}(a_d,b_d)-I_0(a_d,b_d))}=\frac{p^{a_d-1}(1-p)^{b_d-1}}{\mathrm{B}(a_d,b_d)I_{p_0}(a_d,b_d)}
\end{align}
where $I_x(a_d,b_d)$ denotes the regularized incomplete beta function, defined as $I_x(a_d,b_d):=\frac{\mathrm{B}(x;a_d,b_d)}{\mathrm{B}(a_d,b_d)}$, $\mathrm{B}(x;a_d,b_d):=\int_0^x t^{a_d-1}(1-t)^{b_d-1}dt$ is the incomplete beta function, and $\mathrm{B}(a_d,b_d) = \mathrm{B}(1;a_d,b_d)$ is the Beta function. The last equality in \Cref{eq:truncatedBetaDensity} follows from $I_0(a_d,b_d)=0$. The regularized incomplete beta function is the cumulative distribution function of the beta distribution. In brief terms, the truncated Beta distribution on $H_0$ simply normalizes the untruncated Beta distribution on $[0,1]$ to the interval $[0,p_0]$ of $H_0:p\leq p_0$.

Likewise, we pick the truncated Beta distribution under $H_1:p>p_0$, this time normalized to $[p_0,1]$:
\begin{align}\label{eq:designPriorH1OneSidedTesting}
    &p \mid H_1 \sim \mathrm{Beta}(a_d,b_d)_{[p_0,1]}
\end{align}

For the analysis prior $P^a$ under $H_0$ and $H_1$, we also choose truncated Beta priors, with possibly different values $a_a$ and $b_a$, where the subscript signals that the hyperparameters belong to our analysis instead of design prior:
\begin{align}
    &p \mid H_0 \sim \mathrm{Beta}(a_a,b_a)_{[0,p_0]}\label{eq:analysisPriorsOneSidedTestingH0}\\
    &p \mid H_1 \sim \mathrm{Beta}(a_a,b_a)_{[p_0,1]}\label{eq:analysisPriorsOneSidedTestingH1}
\end{align}

\begin{figure}[!htb]
    \centering
    \includegraphics[width=\linewidth]{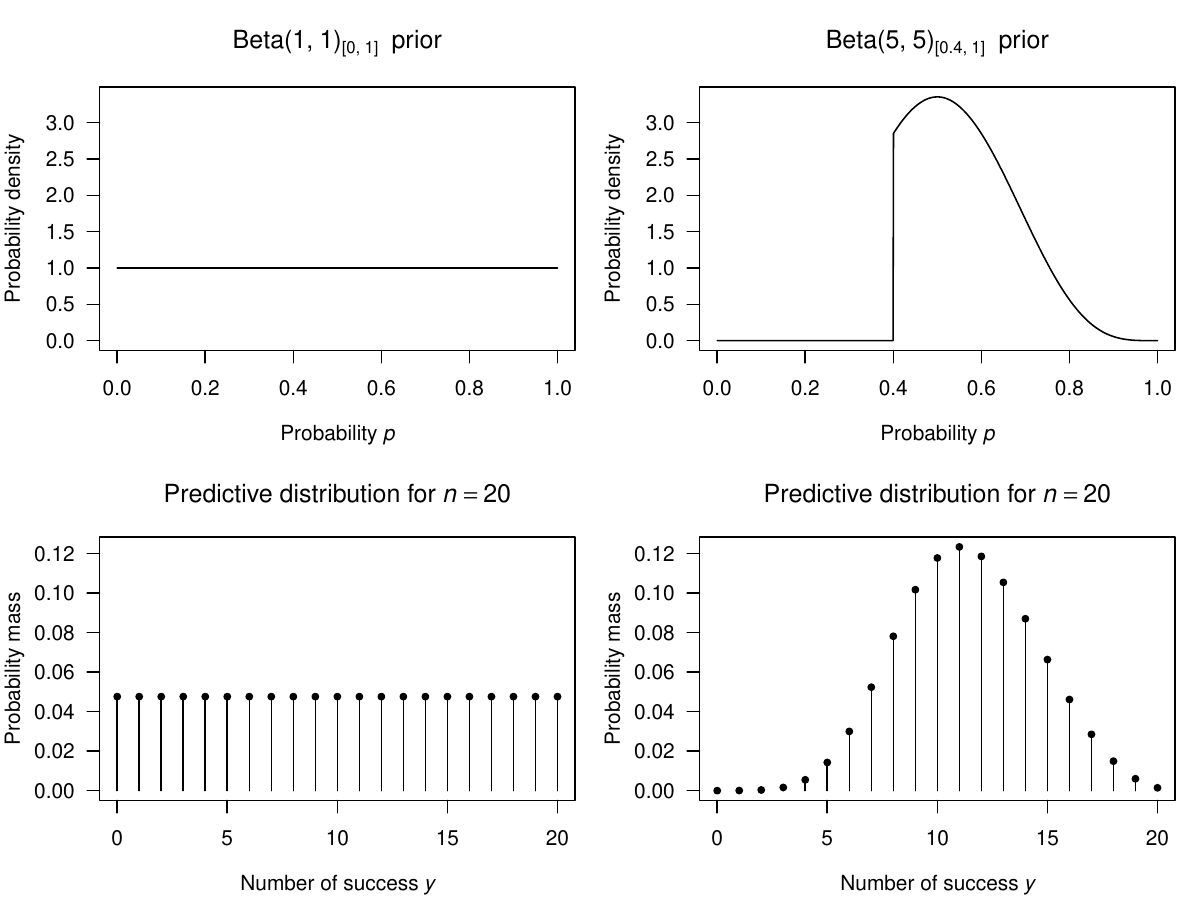}
    \caption{Illustration of two (truncated) beta prior distributions (top) and the corresponding predictive distributions (below).}
    \label{fig:prior-predictive-illustration}
\end{figure}

These priors allow for huge flexibility in incorporating both subjective and objective beliefs about $p$ under $H_0$ respectively $H_1$. For example, a uniform prior is obtained when $a_a=b_a=1$ respectively $a_d=b_d=1$, and other values of the hyperparameters can indicate strong a priori beliefs about the truth of certain parameter regions in $[0,1]$ \citep{Kruschke2015}, see Figure~\ref{fig:prior-predictive-illustration} for an illustration. Also, the Beta distribution has the property that for large $m$ 
\begin{align}\label{eq:freqLimit}
    \mathrm{Beta}(m\cdot a,m\cdot b)\xrightarrow[]{} \mathcal{N}\left(\frac{a}{a+b},\frac{ab}{(a+b)^3}\frac{1}{m}\right)
\end{align}
More precisely, if $Y_m\sim \mathrm{Beta}(m\cdot a, m\cdot b)$, then $\sqrt{m}(Y_m-\frac{a}{a+b})$ converges in distribution to $\mathcal{N}(0,\frac{a}{(a+b)^3})$ as $m$ increases.
Therefore, choosing $\frac{a}{a+b}:=p_0$ and letting $m\rightarrow \infty$ with $a+b$ large enough therefore amounts to using a highly informative prior which converges against a Dirac measure on $p_0$, and which equals a frequentist power calculation in the sense that the Bayesian prior calculates power (or the Bayesian type-I-error rate) under the point-value $p_0$. Importantly, using such a prior reduces $H_0:p\leq p_0$ to the point-null hypothesis $H_0:p=p_0$, and therefore the directional test of $H_0:p\leq p_0$ against $H_1:p>p_0$ reduces to the one-sided test of $H_0:p=p_0$ against $H_1:p>p_0$.

\subsubsection{Derivation of the prior-predictive distribution}
The third step shown in \Cref{fig:overview} consists of deriving the prior-predictive probability mass function $f(y \mid H_0)$ and $f(y \mid H_1)$ under the null and alternative hypothesis. These probability mass functions will be used together with the Bayes factor $\mathrm{BF}_{01}(y)$ later, to compute critical values $y_{crit}$, for which we can state that $\mathrm{BF}_{01}$ passes a given threshold such as $k$. Based on the predictive distributions we can then compute the desired probabilities $P(\mathrm{BF}_{01}(y)<k \mid H_0)$ and $P(\mathrm{BF}_{01}(y)<k \mid H_1)$, compare \Cref{eq:bayesianDesign}. Achieving a specified power will then proceed by selecting the smallest sample size $n\in \mathbb{N}$ for which the predictive probability
$$P(y>y_{crit} \mid H_1)\geq 1-\beta$$
for a specified $\beta$ such as $\beta:=0.2$, where $P(\cdot  \mid H_1)$ denotes the predictive distribution under $H_1$.

Based on the standard Beta prior, marginalizing out $p$ of the binomial likelihood yields the posterior predictive distribution which is Beta-Binomial
\begin{align}
    Y\sim \text{Beta-Binom}(n,a+y,b+n-y)
\end{align}
where $y$ successes have been observed out of $n$ samples. However, here we base the predictive probability mass function on the truncated beta design priors under $H_0$ and $H_1$ to obtain a prior-predictive distribution. This leads to the following predictive probability mass function of the data $Y$ based on a fixed $n$ and a truncated beta prior $p \sim \mathrm{Beta}(a, b)_{[l, u]}$ for $u,l \in [0,1]$ with $l<u$, for details see the Appendix:
\begin{align}\label{eq:predDensity}
        f(y \mid n, a, b, l, u) ={n\choose y}\frac{\mathrm{B}(a+b,b+n-y)[I_u(a+y,b+n-y)-I_l(a+y,b+n-y)]}{\mathrm{B}(a,b)[I_u(a,b)-I_l(a,b)]}
\end{align}

Conditioning on $H_0$ then leads to $f(y \mid n, a, b, l, u, H_0)$, where in the above, we can replace $u$ by $p_0$ and $l$ by $0$. For $H_1$, we obtain $f(y \mid n, a, b, l, u, H_1)$ where we replace $u$ by $1$ and $l$ by $p_0$. Depending on which design prior we use, we further replace $a$ and $b$ by the selected values $a_d$ and $b_d$ in the design prior under $H_0$ in \Cref{eq:designPriorH0OneSidedTesting}, respectively in the design prior under $H_1$ in \Cref{eq:designPriorH1OneSidedTesting}. We stress that the predictive distribution in \Cref{eq:predDensity} makes use of the design priors chosen in advance.



\subsubsection{Derivation of the Bayes factor}
The next step is step 4 in \Cref{fig:overview}, the derivation of the Bayes factor. Therefore, we must choose analysis priors $P^{a}$ under $H_0$ and $H_1$. We opt for the same truncated beta priors here, stressing that it is well possible to choose different hyperparameters $a_a$ and $b_a$ in the analysis than in the design prior $P^d$ (where the hyperparameters were $a_d$ and $b_d$). It is, of course, possible to choose design and analysis prior identically.

For $H_0 \colon p \leq p_0$ versus $H_1 \colon p > p_0$ with truncated Beta priors $p \mid H_0 \sim \mathrm{Beta}(a_a, b_a)_{[0, p_0]}$ and $p \mid H_1 \sim \mathrm{Beta}(a_a, b_a)_{(p_0, 1]}$ we arrive at the Bayes factor
    \begin{align}\label{eq:bayesFactorOneSided}
        \mathrm{BF}_{01}(y) 
        &= \frac{I_{p_0}(a_a + y, b_a + n - y)}{1 - I_{p_0}(a_a + y, b_a + n - y)} \, \frac{1 - I_{p_0}(a_a, b_a)}{I_{p_0}(a_a, b_a)}
    \end{align}
    where a derivation is provided in the Appendix.

\subsubsection{Numerical root-finding}
Based on the Bayes factor, the fifth step in \Cref{fig:overview} now consists of numerical root-finding. In brief terms, the idea is to find a solution to the equation
\begin{align}\label{eq:rootfinding}
    \mathrm{BF}_{01}(y)=k
\end{align}
by numerical means for a fixed sample size $n$. Thus, we use \Cref{eq:bayesFactorOneSided} and numerically find the root of $\mathrm{BF}_{01}(y)-k=0$ via Newton's method as, for example, implemented in the \texttt{uniroot} function in the statistical programming language R \citep{RProgrammingLanguage}. 

\subsubsection{Computation of critical value(s)}
The solution of \Cref{eq:rootfinding} yields a value $y_{crit}$ based on which we can state that for $y>y_{crit}$ we have $\mathrm{BF}_{01}<k$ (e.g. $k=1/10$), that is, evidence for the alternative -- as measured by $\mathrm{BF}_{10}(y)$ -- is at least $1/k=10$. Note that the analysis prior is used to compute the Bayes factor.

\subsubsection{Computation of Bayesian type-I-error rate and power}
Based on the root $y_{crit}$ found in the last step, the seventh step in \Cref{fig:overview} consists of computing the relevant quantities
$P(\mathrm{BF}_{01}(y)<k \mid H_0)$ and $P(\mathrm{BF}_{01}(y)<k \mid H_1)$
where we have
$$P(\mathrm{BF}_{01}(y)<k \mid H_0)=P(y>y_{crit} \mid H_0)$$
and
$$P(\mathrm{BF}_{01}(y)<k \mid H_1)=P(y>y_{crit} \mid H_1)$$
and the former is equal to the Bayesian type-I-error rate, while the latter is the Bayesian analogue to frequentist power, compare \Cref{eq:bayesianDesign}.

\subsubsection{Sample size calculation for the Bayes factor}
The ultimate goal now is to obtain the sample size $n$ for which we can state that $P(\mathrm{BF}_{01}(y)<k \mid H_1)=P(y>y_{crit} \mid H_1)$ exceeds a given threshold, such as $1-\beta$ for $\beta:=0.2$, so we have at least $80\%$ Bayesian power to find at least evidence $1/k$ for $H_1$. It might happen that for a fixed $n$ there is no $y_{crit}$ so that the threshold $\mathrm{BF}_{01}(y)< k$ is fulfilled. Then, increasing $n$ will eventually lead to the situation where a large (or small) enough Bayes factor can be found. This is due to the fact that for a large enough sample size $n$ there must be a number of successes $y$ for which $\mathrm{BF}_{01}(y)<k$ holds due to the asymptotic properties of the Bayes factor \citep{Kleijn2022}.

\subsection{Two-sided hypothesis test}

Now, the second option is to use a two-sided hypothesis test of $H_0:p=p_0$ versus $H_1:p\neq p_0$. 

\subsubsection{Design and analysis priors}
Based on the same assumption $Y \mid p \sim \mathrm{Bin}(n, p)$, a Dirac measure in $p_0$ is chosen as the analysis prior under $H_0$, that is,
$$p \mid H_0 \sim \delta_{p_0}$$
where as under $H_1$, a Beta prior is selected:
$$p \mid H_1 \sim \mathrm{Beta}(a_a,b_a)$$
with hyperparameters $a_a,b_a$.

The design priors are chosen as follows: Under $H_1$, the lower truncation point is $0$, and the upper truncation point $1$. The design prior under $H_1$ therefore becomes a normal Beta prior $\mathrm{Beta}(a_d,b_d)$ with hyperparameters $a_d,b_d$. Under $H_0$, a Dirac measure is chosen in $p_0$ again.

\subsubsection{Derivation of the prior-predictive}
From \Cref{eq:predDensity} we obtain now the prior-predictive distribution by setting $l=0$ and $u=1$ under $H_1$. Under $H_0$, the prior-predictive probability mass function reduces to
$$f(y \mid n)=\mathrm{Bin}(y \mid n,p_0)$$
Under $H_1$, we can use the design prior given in \Cref{eq:predDensity} with $l=0$ and $u=1$ and appropriate hyperparameters $a_d$ and $b_d$ plugged in for $a$ and $b$.

\subsubsection{Derivation of the Bayes factor}
    
For the point null test of $H_0 \colon p = p_0$ versus $H_1 \colon p \neq p_0$ with prior $p \mid H_1 \sim \mathrm{Beta}(a, b)$, the Bayes factor results in
    \begin{equation}\label{eq:twoSidedBayesFactor}
        \mathrm{BF}_{01}(y) = p_0^y \, (1 - p_0)^{n - y} \, \frac{\mathrm{B}(a , b)}{\mathrm{B}(a + y, b + n - y)}
    \end{equation}
where a derivation is provided in the Appendix.

\section{Examples}\label{sec:examples}

\subsection{Single-arm phase II proof of concept trial with binary endpoint}\label{subsec:phaseII}

In this subsection, we adopt an example of a clinical trial discussed in \cite{KelterSchnurr2024} in the context of sequential design. There, a single-arm phase IIA study to demonstrate the efficacy of a novel drug or therapy is considered and the primary endpoint is binary. The primary objective of the study was to assess the efficacy of a combination therapy as front-line
treatment in patients with advanced nonsmall cell lung cancer. The study involved
the combination of a vascular endothelial growth factor antibody plus an epidermal
growth factor receptor tyrosine kinase inhibitor. The primary endpoint is the clinical
response rate, that is, the rate of complete response and partial response combined,
to the new treatment. The hypotheses of interest are given as
$$H_0:p\leq p_0 \text{ versus } H_1:p>p_1$$
The current standard treatment yields a response rate of $\approx$ 20\% , so we have
$p_0=0.2$. The target response rate of the new regimen is 40\%, so $p_1 = 0.4$. We stress that somewhat unrealistic, the parameter interval $(0.2,0.4]$ is somehow excluded from the hypotheses considered, a choice that is sometimes seen in phase IIA studies, compare \cite{Lee2008}. While such a distance between $H_0$ and $H_1$ allows for an easier separation between both hypotheses via statistical testing,  we refrain from doing so and instead select a flat prior $\mathrm{Beta}_{[0,p_0]}(1,1)$ with $a_d=b_d=1$ to stay neutral first. Later, we also opt for the more realistic choice to center a truncated Beta design prior on $p_1=0.4$, so we select $0.4$ for the mode of the design prior, details follow below. Under $H_0:p\leq p_0$, we select a flat Beta design prior, and we formally test
$$H_0:p\leq 0.2 \text{ versus } H_1:p>0.2$$
here. For the analysis priors, we also select flat truncated Beta priors both under $H_0$ and $H_1$, that is, $a_a=b_a=1$ in \Cref{eq:analysisPriorsOneSidedTestingH0} and \Cref{eq:analysisPriorsOneSidedTestingH1}. This choice of design and analysis priors intuitively seems objective.

\begin{figure}[!htb]
    \centering

    \begin{subfigure}[b]{0.49\textwidth}
    \includegraphics[width=\linewidth]{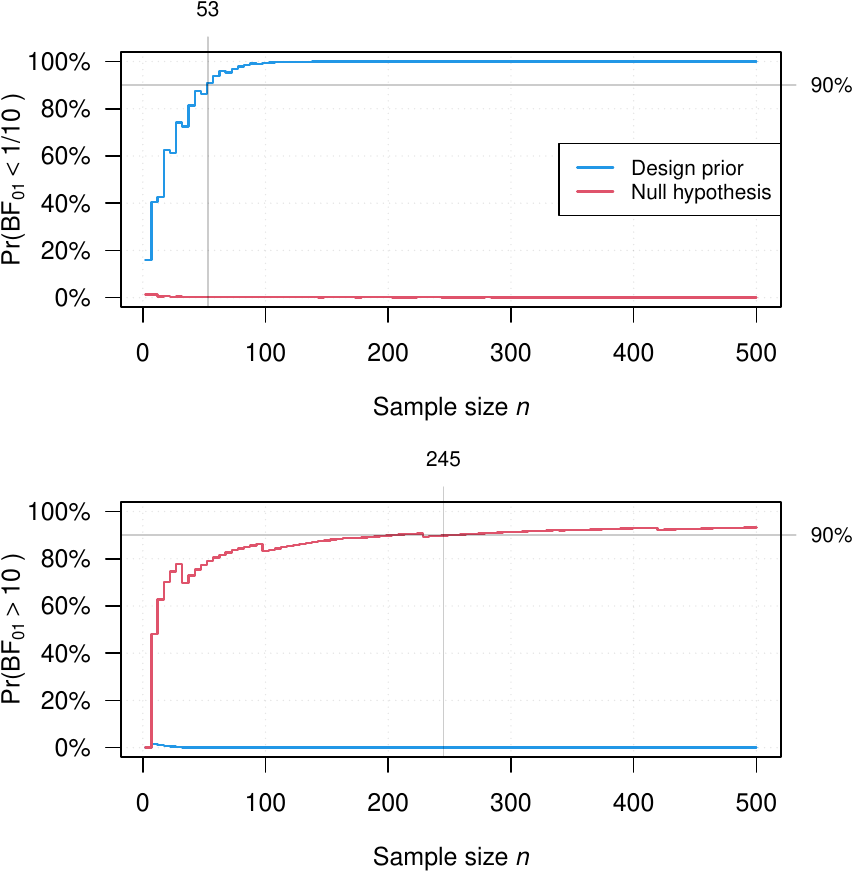}
    \caption{Evidence threshold $k=1/10$}
    \label{fig:ex1_4}
    \end{subfigure}
    \hfill
    \begin{subfigure}[b]{0.49\textwidth}
        \centering
        \includegraphics[width=\textwidth]{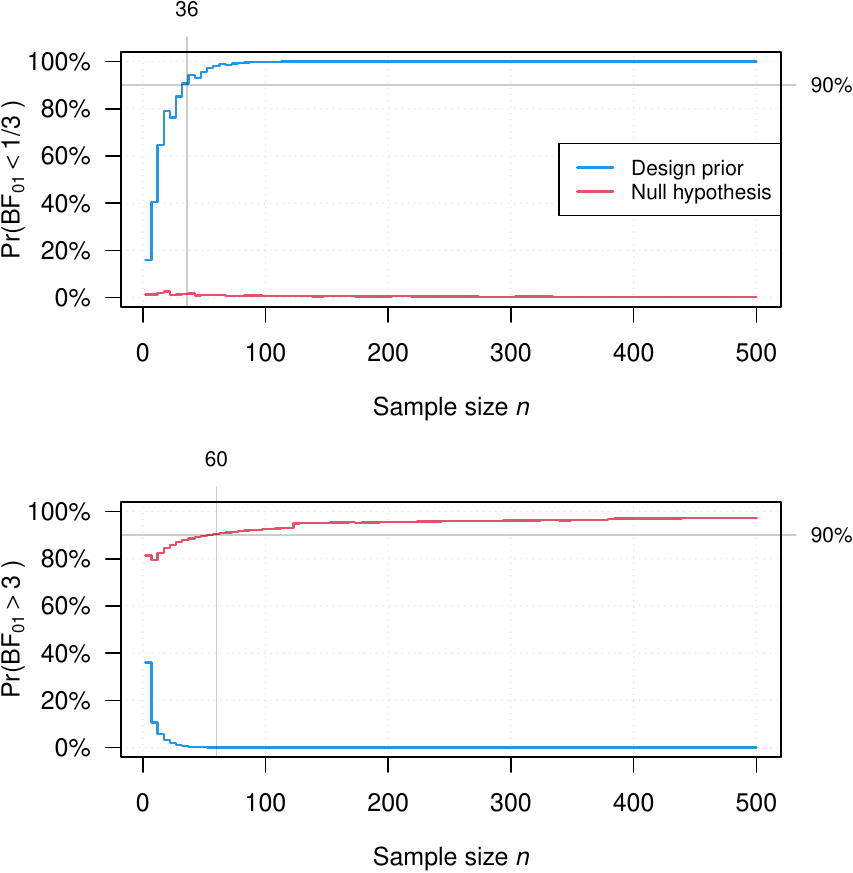}
        \caption{Evidence threshold $k=1/3$}
        \label{fig:ex1_3}
    \end{subfigure}
    
    \caption{Top plot shows the Bayesian power $P(\mathrm{BF}_{01}(y)<k|H_1)$ under $H_1$ as the solid blue line, and the Bayesian type-I-error rate $P(\mathrm{BF}_{01}(y)<k|H_0)$ under $H_0$ as the solid red line. The bottom plot shows the Bayesian power to obtain compelling evidence in favour of $H_0$, $P(\mathrm{BF}_{01}(y)>1/k|H_0)$. In all cases, a point design prior at $p_1=0.4$ is assumed under $H_1$, while the design prior under $H_0$ as well as the analysis priors under $H_0$ and $H_1$ are uniform.}
    \label{fig:ex1_34}
\end{figure}

\begin{figure}
    \centering
    \includegraphics[width=1.0\linewidth]{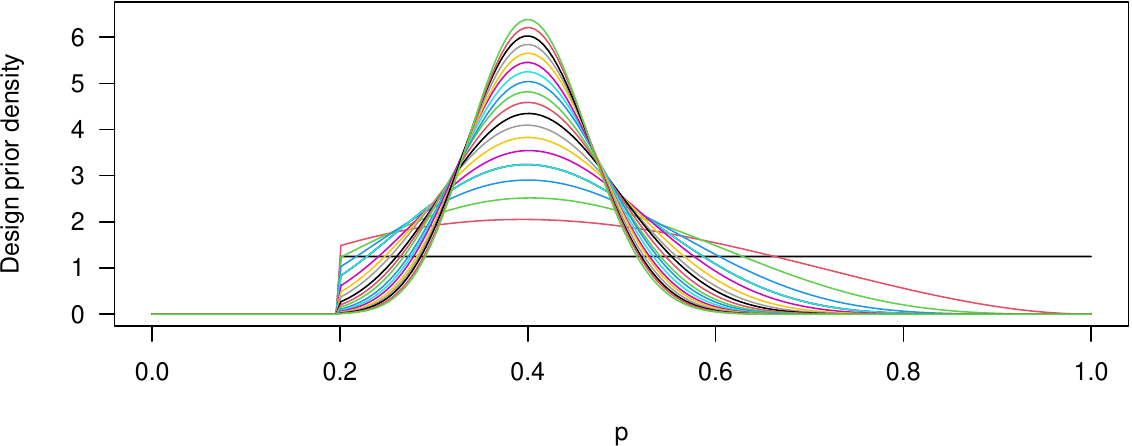}
    \caption{Truncated $\mathrm{Beta}_{[0.2,1]}(a_d,b_d)$ design priors centered at the mode $p_1=0.4$ for increasingly informative hyperparameters $a_d$ and $b_d$. The relationship between $a_d$ and $b_d$ is given in \Cref{eq:ex1_informativeDesignPrior_ad}.}
    \label{fig:ex_1_informativeDesignPriors}
\end{figure}

In \cite{KelterSchnurr2024}, a sequential design was used and calibrated via a Monte Carlo simulation to attain the boundaries $\alpha\leq 0.1$ and $\beta \leq 0.1$ for the Bayesian type-I-error rate and power. This choice is widely adopted in phase II trials \citep{Lee2008,Zhou2017,Berry2011}, so we opt for these boundaries in the context of our Bayesian design requirements given in \Cref{eq:bayesianDesign}. However, we do not use a sequential but fixed sample size design, which allows for \textit{much} easier conduct of the trial, possibly at the cost of a less efficient design. Still, sequential designs are logistically much more challenging, and additionally troubled by the need to calibrate a Monte Carlo simulation and attain sufficiently small Monte Carlo standard errors \citep{Zhou2023}. Thus, we need to calculate a sample size so that
$$P(\mathrm{BF}_{01}(y) < k \mid H_0)\leq 0.1 \hspace{1cm} \text{and} \hspace{1cm} P(\mathrm{BF}_{01}(y) < k \mid H_1)> 0.9$$
where we select the threshold $k=1/10$ for strong evidence. 
\begin{figure}[!tb]
    \centering
    
     \begin{subfigure}[b]{0.49\textwidth}
         \centering
         \includegraphics[width=\textwidth]{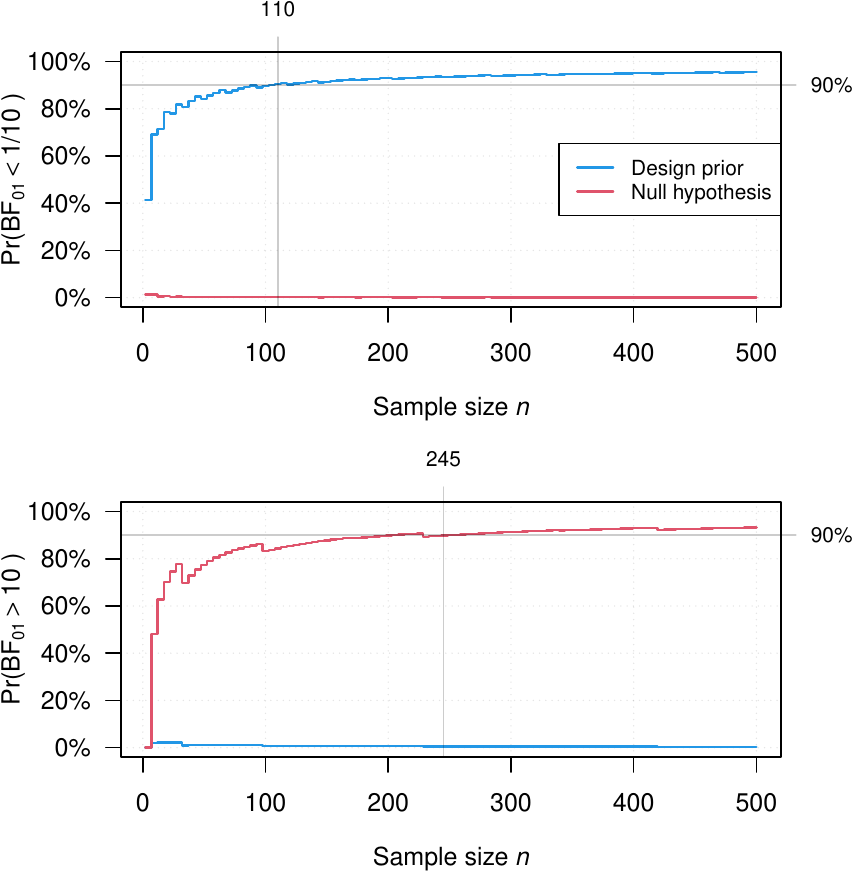}
         \caption{Evidence threshold $k=1/10$}
         \label{fig:ex1_1}
     \end{subfigure}
     \hfill
     \begin{subfigure}[b]{0.49\textwidth}
         \centering
         \includegraphics[width=\textwidth]{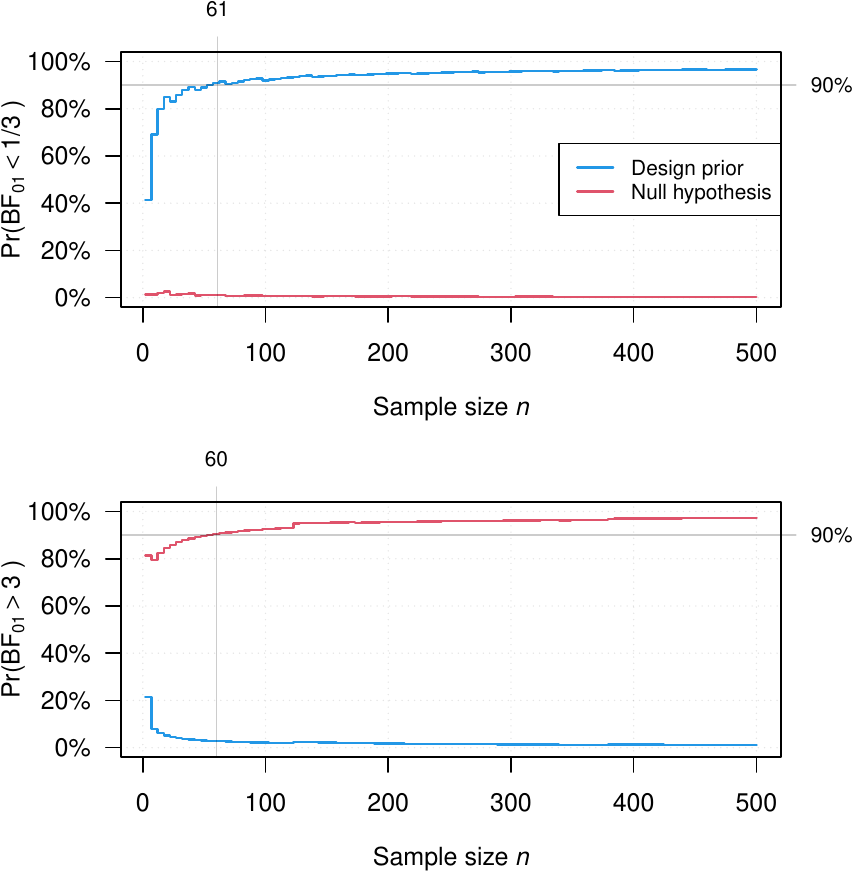}
         \caption{Evidence threshold $k=1/3$}
         \label{fig:ex1_2}
     \end{subfigure}
     
    \caption{Bayesian power and type-I-error rate for the single-arm phase II proof of concept trial with binary endpoint. Top plots in each subfigure show the Bayesian power $P(\mathrm{BF}_{01}(y)<k \mid H_1)$ under $H_1$ as the solid blue line, and the Bayesian type-I-error rate $P(\mathrm{BF}_{01}(y)<k \mid H_0)$ under $H_0$ as the solid red line. The bottom plots in each subfigure show the Bayesian power to obtain compelling evidence in favour of $H_0$, $P(\mathrm{BF}_{01}(y)>1/k \mid H_0)$. Flat design and analysis priors are assumed in all cases.}
    \label{fig:ex1}
\end{figure}
Application of the root-finding approach shown in \Cref{fig:overview} then yields a sample size of $n = 110$. This sample size leads to a power of $90.05\%$ (under $H_1$) under the truncated design beta design prior $\mathrm{Beta}(a = 1, b = 1)_{[0.2, 1]}$ , while under the null (i.e., a uniform prior from 0 to 0.2) the Bayesian type-I-error rate is $\alpha = 0.16\%$. The top plot in Figure~\ref{fig:ex1_1} shows the corresponding Bayesian power and type-I-error curves.\footnote{In the binomial model, it happens that power and type I error viewed as a function of the sample size $n$ show ``oscillation'' or ``zig-zag'' behaviours as visible in Figure~\ref{fig:ex1}-\ref{fig:ex2_res_twoSided_moderate}. That is, due to the discreteness of the data, it can happen that the power (type I error rate) increases (decreases) but then decreases (increases) again when the sample size is increased. These behaviours are not limited to Bayes factors but occur for many other methods in the binomial model, see e.g. chapter 17 in \citet{Julious2023} for illustration of oscillating power of binomial tests, or section 2 in \citet{Brown2001} for oscillating coverage rates of confidence intervals for a binomial proportion. For this reason, it is important to verify that a calculated sample size truly ensures a power (type I error rate) above (below) the target power (type I error rate) for any greater sample size. In the \texttt{bfpwr} package this is implemented via an algorithm that terminates the numerical search only if power (type I error) does not drop below (increase above) the target for at least 10 additional observations, otherwise the numerical search is continued.} The blue solid line is the Bayesian power, while the red solid line is the associated Bayesian type-I-error rate, compare \Cref{eq:bayesianDesign}. 

Finally, the type-I-error rate and power related to $n = 110$ under the point values $p_0 = 0.2$ and $p_1 = 0.4$ are given by $2.47\%$ and $99.63\%$, respectively. Hence, the design is calibrated also from a fully frequentist perspective -- which uses Bayes factors as test statistics, a concept originally proposed by \cite{Good1983a}. 

Now, the bottom plot in Figure~\ref{fig:ex1_1} additionally shows the probability to obtain compelling evidence in form of a Bayes factor larger than $1/k=10$ in favour of the null hypothesis $H_0:p\leq 0.2$. This probability can be denoted as the \textit{Bayesian power to obtain compelling evidence for $H_0$}.  
Thus, for $n=245$ patients, we have at least 90\% power to accept $H_0$ based on strong evidence with $1/k=10$. Note that such a perspective on Bayesian power is not available in frequentist testing, and also not formally required according to \Cref{eq:bayesianDesign}. However, in practice, stopping a phase II a trial when a drug is inefficient is highly desirable \citep{Berry2011,Lee2008,Zhou2017,KelterSchnurr2024}. One reason is that due to the possibly large number of potentially effective agents, filtering promising ones from ineffective drugs is important. This is one primary reason to consider sequential designs after all, as they allow early stopping for futility, see for example \cite{Zhou2017} and \cite{KelterSchnurr2024}. In the calibrated design shown in \Cref{fig:ex1_1}, we can be quite certain -- with at least $90\%$ probability -- that we will stop for futility based on $n=245$ patients.

The sample sizes $n=110$ and $n=245$ seem prohibitively large for a phase II trial. Thus, \Cref{fig:ex1_2} demonstrates the effect of using the same design and analysis priors with a more liberal threshold $k=1/3$, which amounts to only moderate evidence \citep{Jeffreys1939}. This is an assumption which is readily justified in the context of a single-arm phase II trial. As shown in the upper plot of \Cref{fig:ex1_2}, now $n=61$ patients suffice to achieve at least 90\% Bayesian power, and the probability to obtain compelling evidence in favour of $H_0$ -- as shown in the bottom plot of \Cref{fig:ex1_2} -- reaches 90\% already after $n=60$ patients. We see that based on the more realistic threshold $k=1/3$ in a phase II trial, the sample sizes required for carrying out such a study become realistic.


The advantage of the two sample size calculations above are their realism: The truncated design priors include also parameter values very close to the boundary $p_0$ between $H_0$ and $H_1$, and thus the calculated sample sizes become large because separation between $H_0$ and $H_1$ becomes naturally more difficult. The advantage of the realism is the price paid in sample size required, and a practically possible solution to this aspect is to consider the frequentist limit as described in \Cref{eq:freqLimit}. 


Thus, we reduce the truncated Beta design prior to a point-prior in $p_1=0.4$ next. Sample sizes under $H_1$ are thus calculated based on a single parameter value without taking into account its uncertainty. Sample size calculation under $H_0$ proceeds as before in \Cref{fig:ex1_2}. The bottom part of \Cref{fig:ex1_3} therefore yields identical results like the one in \Cref{fig:ex1_1}, and after $n=60$ patients we can state with at least 90\% probability that the drug is ineffective if $H_0:p\leq 0.2$ holds. The top part in \Cref{fig:ex1_3} now shows that under the point-prior with all probability mass on $p_1=0.4$ we arrive at $n=36$ patients required to yield a calibrated Bayesian design based on the Bayes factor. Note that under $H_0$ -- see the bottom plot in \Cref{fig:ex1_3} -- nothing changes, because the Bayesian power to find compelling evidence for the null is still averaged under the truncated Beta prior on $H_0$, that is, $p\mid H_0 \sim \mathrm{Beta}(1,1)_{[0,0.2]}$.

In sum, shifting the evidence threshold to $k=1/3$ yields achievable and realistic sample sizes in terms of calibrating Bayesian power and type-I-error rate, as shown above. Additionally, the sample sizes required to calibrate the probability to find compelling evidence for the null hypothesis are also practically attainable.

However, by now all sample size calculations were based on a flat design prior $\mathrm{Beta}_{(0.2,1]}(1,1)$ with $a_d=b_d=1$ under $H_1$, compare \Cref{eq:designPriorH1OneSidedTesting}, and $\mathrm{Beta}_{(0,0.2]}(1,1)$ with $a_d=b_d=1$ under $H_0$, compare \Cref{eq:designPriorH0OneSidedTesting}. A different choice is given by using a design prior that is centered at some $p_1>p_0$, so under $H_1$ we have the expectation that the drug is effective in some way.\footnote{Importantly, we must not necessarily believe a priori that such an effectiveness is most probable without any observed data. We can also just interpret such a centered prior as the design prior which expresses the sample size requirements for a minimally relevant effect size of interest. If, e.g. we center the prior at $p_1=0.4$, we can argue that the resulting sample sizes based on the root-finding approach are most realistic if the true effect size is somewhere around $p_1=0.4$. If the drug is more effective, the calculated sample sizes will be too large. If the drug is less effective, they will be too small. The latter two aspects depend in turn on how strongly centered the design prior is around $p_1=0.4$.}Therefore, we consider three scenarios: First, we could center the mean of the Beta distribution at $p_1$. Second, we could center the mode of the Beta distribution at $p_1$. Third, we could center both at $p_1$. As shown in the Appendix, the third option is possible only if $p=0.5$, which is not helpful. Therefore, we need to decide between mean- and mode-centering at some $p_1 >p_0$. We pick mode-centering, primarily because this aligns with the notion of a frequentist power analysis where we are interested in a specific point $p_1$ under the alternative hypothesis $H_1$. Thus, mode-centering allows to specify which is the most probable point under $H_1$ according to our assumptions in the design stage.

\Cref{fig:ex_1_informativeDesignPriors} shows several truncated $\mathrm{Beta}_{[0.2,1]}(a_d,b_d)$ design priors centered at the mode $p_1=0.4$ for increasingly informative hyperparameters $a_d$ and $b_d$. As the mode of the truncated Beta distribution is given as $\frac{a_d-1}{(a_d+b_d-2)}$, it is possible to obtain from $\frac{a_d-1}{(a_d+b_d-2)}=p_1$ for some $p_1 \in (0,1)$ that
\begin{align}\label{eq:ex1_informativeDesignPrior_ad}
    a_d = \frac{p_1(b_d - 2) + 1}{1 - p_1}
\end{align}
for details see the Appendix. Therefore, for each chosen value of $b_d$, selecting $a_d$ as in \Cref{eq:ex1_informativeDesignPrior_ad} yields a truncated Beta design prior centered at $p_1$. For $p_1=0.4$, one obtains the values shown in \Cref{tab:ex1_full}. The parameters at the top are less informative, and the parameters at the bottom are the most informative. This can also be interpreted as follows: For example, $a_d=5$ and $b_d=7$ implies that the informativity of our prior equals the information which is obtained from observing $a_d+b_d=12$ patients before conducting the trial, a well-known fact from the beta-binomial model \citep{Kruschke2015a}. Out of these 12 patients, we assume that $5$ successes and $5$ failures have been observed for $a_d=5$ and $b_d=7$. This information is shown in the third column in \Cref{tab:ex1_full}. From top to bottom, this information increases from $2$ to $62$ patients for $a_d=25$ and $b_d=37$. Thus, the most informative prior centered at $p_1=0.4$ yields as much information as having observed $62$ patients already, which is very large in the context of a phase II trial \citep{Berry2011}. Note that using a prior which assumes that we have information that resembles $a_d=25$ observed successes and $b_d=37$ failures might be unrealistically informative in the context of a phase II trial, depending on the context.

\begin{table}[!p]
\centering
\resizebox{0.975\textwidth}{!}{
\begin{tabular}{c|c|c|c|c|c|c|c}
  \toprule
$a_d$ & $b_d$ & $a_d+b_d$ & Sample size $n$ & $P(\mathrm{BF}_{01}(y)<k|H_1)$ & $P(\mathrm{BF}_{01}(y)<k|H_0)$ & FP & FT1E \\ 
  \midrule
  \multicolumn{8}{l}{Evidence threshold $k = 1/10$} \\
  \cmidrule(lr){1-8}
    1.0 & 1 & 2.0 & 110 & 90.05 & 0.16 & 99.63 & 2.47 \\ 
  2.3 & 3 & 5.3 & 196 & 90.12 & 0.24 & 100.00 & 2.44 \\ 
  3.7 & 5 & 8.7 & 183 & 90.18 & 0.35 & 100.00 & 2.47 \\ 
  5.0 & 7 & 12.0 & 170 & 90.15 & 0.46 & 99.99 & 2.48 \\ 
  6.3 & 9 & 15.3 & 157 & 90.25 & 0.56 & 99.98 & 2.48 \\ 
  7.7 & 11 & 18.7 & 132 & 90.31 & 0.75 & 99.92 & 2.70 \\ 
  9.0 & 13 & 22.0 & 123 & 90.25 & 0.80 & 99.84 & 2.56 \\ 
  10.3 & 15 & 25.3 & 111 & 90.24 & 0.99 & 99.70 & 2.79 \\ 
  11.7 & 17 & 28.7 & 106 & 90.45 & 0.97 & 99.55 & 2.53 \\ 
  13.0 & 19 & 32.0 & 98 & 90.35 & 1.12 & 99.30 & 2.66 \\ 
  14.3 & 21 & 35.3 & 94 & 90.49 & 1.22 & 99.14 & 2.72 \\ 
  15.7 & 23 & 38.7 & 86 & 90.36 & 1.37 & 98.67 & 2.84 \\ 
  17.0 & 25 & 42.0 & 85 & 90.06 & 1.23 & 98.37 & 2.47 \\ 
  18.3 & 27 & 45.3 & 85 & 90.45 & 1.27 & 98.37 & 2.47 \\ 
  19.7 & 29 & 48.7 & 81 & 90.46 & 1.35 & 97.98 & 2.51 \\ 
  21.0 & 31 & 52.0 & 81 & 90.78 & 1.39 & 97.98 & 2.51 \\ 
  22.3 & 33 & 55.3 & 77 & 90.39 & 1.46 & 97.50 & 2.54 \\ 
  23.7 & 35 & 58.7 & 77 & 90.87 & 1.50 & 97.50 & 2.54 \\ 
  25.0 & 37 & 62.0 & 73 & 90.33 & 1.57 & 96.89 & 2.57 \\ 
  \midrule
  \multicolumn{8}{l}{Evidence threshold $k = 1/3$} \\
  \cmidrule(lr){1-8}
  1.0 & 1 & 2.0 & 61 & 90.49 & 0.94 & 98.24 & 8.79 \\ 
  2.3 & 3 & 5.3 & 108 & 90.37 & 1.24 & 99.92 & 8.09 \\ 
  3.7 & 5 & 8.7 & 112 & 90.38 & 1.61 & 99.94 & 7.79 \\ 
  5.0 & 7 & 12.0 & 99 & 90.17 & 2.13 & 99.85 & 7.92 \\ 
  6.3 & 9 & 15.3 & 90 & 90.05 & 2.48 & 99.71 & 7.70 \\ 
  7.7 & 11 & 18.7 & 82 & 90.38 & 3.12 & 99.54 & 8.29 \\ 
  9.0 & 13 & 22.0 & 74 & 90.58 & 3.82 & 99.29 & 8.92 \\ 
  10.3 & 15 & 25.3 & 73 & 90.59 & 3.62 & 99.10 & 7.95 \\ 
  11.7 & 17 & 28.7 & 65 & 90.44 & 4.27 & 98.60 & 8.50 \\ 
  13.0 & 19 & 32.0 & 61 & 90.53 & 4.73 & 98.24 & 8.79 \\ 
  14.3 & 21 & 35.3 & 60 & 90.17 & 4.29 & 97.79 & 7.72 \\ 
  15.7 & 23 & 38.7 & 60 & 90.65 & 4.45 & 97.79 & 7.72 \\ 
  17.0 & 25 & 42.0 & 56 & 90.42 & 4.81 & 97.22 & 7.95 \\ 
  18.3 & 27 & 45.3 & 56 & 90.81 & 4.94 & 97.22 & 7.95 \\ 
  19.7 & 29 & 48.7 & 52 & 90.37 & 5.29 & 96.50 & 8.17 \\ 
  21.0 & 31 & 52.0 & 52 & 90.69 & 5.41 & 96.50 & 8.17 \\ 
  22.3 & 33 & 55.3 & 52 & 90.99 & 5.51 & 96.50 & 8.17 \\ 
  23.7 & 35 & 58.7 & 48 & 90.30 & 5.84 & 95.58 & 8.38 \\ 
  25.0 & 37 & 62.0 & 48 & 90.54 & 5.93 & 95.58 & 8.38 \\ 
  \bottomrule
\end{tabular}
}
\caption{Increasingly informative hyperparameters $a_d$ and $b_d$, rounded to three digits, for the truncated $\mathrm{Beta}_{[0.2,1]}(a_d,b_d)$ design priors centered at the mode $p_1=0.4$ in the phase II trial example. The fourth column shows the resulting sample size $n$ to achieve $90\%$ Bayesian power $P(\mathrm{BF}_{01}(y)<k|H_1)$. The fifth column is the Bayesian power, the sixth column the Bayesian type-I-error rate, FP the frequentist power evaluated at $p_1=0.4$ based on \Cref{eq:freqLimit}, and FT1E the frequentist type-I-error rate. All probabilities are given as percentages.}
    \label{tab:ex1_full}
\end{table}

The fourth column in \Cref{tab:ex1_full} shows the resulting sample sizes obtained from the root-finding approach. We see that sample size increases first, and then starts decreasing once a given amount of informativity has been reached. This is to be expected, because the Bayesian sample size is an average based on all values in $H_1:p>0.2$. Sample sizes become large, in particular, for parameter values close to $p_0=0.2$. For example, the required sample size based on $p_1=0.21$ will be huge, because it is difficult for the Bayes factor to separate between $H_0$ and $H_1$ then. \Cref{fig:ex_1_informativeDesignPriors} shows that for increasing informativity, the probability of these values decreases -- compare the decreasing progression of density values at parameters slightly larger than $p_0=0.2$ -- so one would expect decreasing sample sizes of the root-finding approach for increasing informativity. However, small sample sizes are achieved in sample size calculations, in particular, for large probabilities close to $1$. For increasing informativity of the $p_1=0.4$ mode-centered design prior, however, the probability of these values decreases, compare the decreasing progression of lines in \Cref{fig:ex_1_informativeDesignPriors} in the region $[0.55,1]$. Thus, as informativity increases according to \Cref{tab:ex1_full}, the resulting sample size is a balance of (i) reduction of sample size, because small success probabilities close to $p_0$ are weighed less, and (ii) increase of sample size, because large success probabilities close to $1$ are weighed less. For increasing informativity, aspect (i) dominates the process and sample sizes start decreasing more and more eventually.

For example, for $b_d=37$, we obtain $a_d=25$, so $62$ patients have already been observed in the sense of the informativity of the design prior. In such an extreme case, sample size required for achieving $90\%$ Bayesian power under $H_1$ with evidence threshold $k = 1/10$ drops to even $n=73$ patients. Note that for $k = 1/10$ the Bayesian type-I-error rate is less than $5\%$ for all choices of the hyperparameters $a_d$ and $b_d$ -- see column $P(\mathrm{BF}_{01}(y)<k|H_0)$ in \Cref{tab:ex1_full} -- so formally \Cref{eq:bayesianDesign} holds and the designs are all calibrated. From a calibration point-of-view the prior with $a_d+b_d=62$ could thus be used, although possibly considered extreme by some. This prior is shown as the top solid line of all lines in \Cref{fig:ex_1_informativeDesignPriors}. Note that a priori it nearly excludes probabilities $p>0.6$ and close to $p_0=0.2$. In contrast, the flat design prior with $a_d=b_d=1$ is shown as the solid black line in \Cref{fig:ex_1_informativeDesignPriors}, and yields the $n=110$ patients shown in \Cref{fig:ex1_1}.

An interesting learning from \Cref{tab:ex1_full} is that to obtain the extremely small sample size of $n=36$ patients of the frequentist power analysis given in \Cref{fig:ex1_3}, one must choose a design prior that is even much more informative than pretending to have observed $62$ patients already. For the latter, we arrive at $n=73$ Patients. Based on our root-finding approach, using even huge hyperparameters such as $b_d=10000$ and $a_d=6667$ -- centered at $p_1=0.4$ -- yields sample sizes such as $n=53$ under $H_1$ to achieve $90\%$ Bayesian power. This demonstrates that a Bayesian, who is using \textit{drastically} \textit{subjective} prior beliefs, is still far away from the small sample size a frequentist is calculating.\footnote{A prior based on $16667$ already observed samples with more than $6000$ successes would render any subsequent phase II trial with $n=53$ patients entirely obsolete.} This can be seen by modification of the results in \Cref{fig:ex1_3} to use the evidence threshold $k=1/10$ again, which is the one used in the bottom part of \Cref{tab:ex1_full}. The results are shown in \Cref{fig:ex1_4}.

The top plot in \Cref{fig:ex1_4} demonstrates that $n=53$ samples are required in the power analysis where a point prior is used at $p_1=0.4$ -- compare \Cref{eq:freqLimit} -- and the results of this frequentist power analysis based on $k=1/10$ are equal to a Bayesian analysis which assumes $a_d+b_d=16667$ patients have already been observed, with $a_d=10000$ failures and $b_d=6667$ successes. Termed otherwise, a frequentist power analysis could be seen with suspicion by almost all Bayesians, whether they choose an objective or subjective prior. From a Bayesian perspective, the required sample size obtained in \Cref{fig:ex1_3} is too optimistically calculated independently of the beliefs about the parameter. However, a frequentist could object that beliefs about the parameter under $H_1$ are irrelevant for power calculations and the sample size should be calculated based on a minimally relevant effect size. Still, a Bayesian could interpret the mode-centered prior under $H_1$ exactly as centered around such a minimally relevant effect, instead of relying on an interpretation in terms of belief.

We close this example by noting that in \Cref{tab:ex1_full}, the frequentist type-I-error rate is always larger than the Bayesian averaged type-I-error rate $P(\mathrm{BF}_{01}(y)<k|H_0)$, which is the price paid by the frequentist analysis to achieve larger power FP compared to the Bayesian powers $P(\mathrm{BF}_{01}(y)<k|H_1)$.

Also, the bottom part of \Cref{tab:ex1_full} shows that when shifting to the evidence threshold $k=1/3$, the required sample sizes range from $48$ to $112$. Note that Bayesian designs with $a_d+b_d>45.33$ yield a Bayesian type-I-error rate of $> 5\%$ and are not calibrated anymore. \Cref{tab:ex1_full} shows that shifting to more liberal evidence threshold $k=1/3$ comes at the price that the smallest sample size which is produced by a calibrated Bayesian design according to \Cref{eq:bayesianDesign} is $n=56$, which still is a reasonable sample size for a phase II proof-of-concept trial that aims at demonstrating the efficacy of a novel drug.

\subsection{Therapeutic touch experiment of Rosa et al.}

In the second example, we revisit a popular experiment reported by \cite{Rosa1998}. Emily Rosa published this study at age nine with the help of her parents, after seeing a television documentary in which so-called therapeutic touch practitioners argued they could sense the human energy field of a person. She tested this supposed ability by letting different practitioners raise their hands through a screen and holding her own hand close to one of the hands. Practitioners then needed to select which of their hands senses Rosa's human energy field, and Rosa recorded the correct and incorrect decisions from therapeutic touch practitioners. According to \cite{Rosa1998}, in the initial trial, the subjects stated the correct location of the investigator's hand in 70 (47\%) of 150 tries. 

\subsubsection{Priors and hypothesis tests}
Two perspectives on a statistical test in the therapeutic touch experiment are possible.
\begin{enumerate}
    \item[$\blacktriangleright$]{First, one could argue to test $H_0:p=\frac{1}{2}$ versus $H_1:p\neq \frac{1}{2}$, because it is expected that the proportion of correct decisions by therapeutic touch practitioners should, under absence of any ability to sense a human energy field, be equal to the toss of a fair coin, that is, $p=\frac{1}{2}$.}
    \item[$\blacktriangleright$]{Second, one could argue that the ability to sense a human energy field exists only, if therapeutic touch practitioners perform better than the toss of a fair coin. That is, the probability of correct decisions should be larger than $\frac{1}{2}$, which leads to the test of $H_0:p\leq \frac{1}{2}$ against $H_1:p>\frac{1}{2}$.}
\end{enumerate}
Both tests can be carried out with the Bayes factors derived in \Cref{sec:rootfinding}. For the two-sided test, the Bayes factor in \Cref{eq:twoSidedBayesFactor} is used, whereas for the directional test, the Bayes factor in \Cref{eq:bayesFactorOneSided} is employed.


\begin{figure}
     \centering
     \begin{subfigure}[b]{0.49\textwidth}
         \centering
         \includegraphics[width=\textwidth]{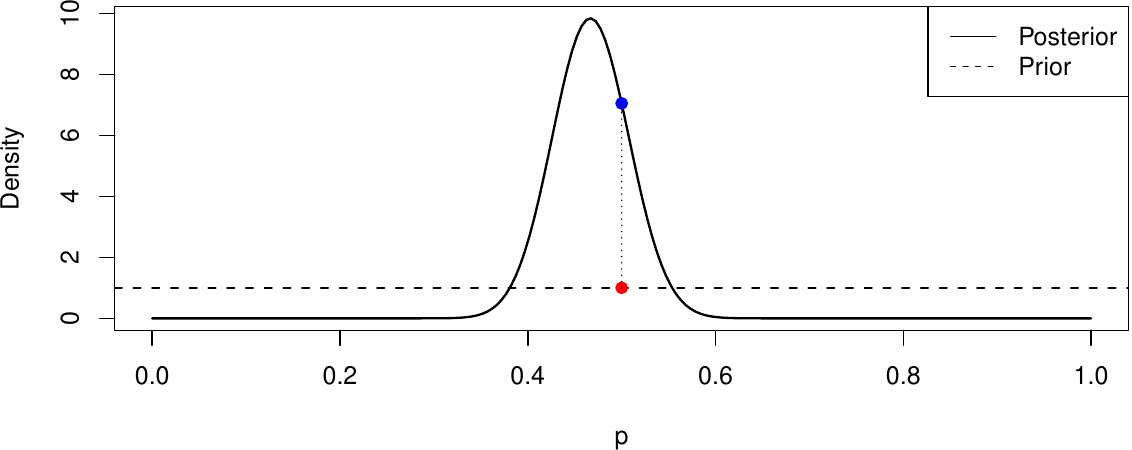}
         \caption{Prior and posterior under $H_1:p\neq \frac{1}{2}$ for the two-sided test of $H_0:p=\frac{1}{2}$ against $H_1:p\neq \frac{1}{2}$ in the therapeutic touch experiment.}
         \label{fig:ex2_1_a}
     \end{subfigure}
     \hfill
     \begin{subfigure}[b]{0.49\textwidth}
         \centering
         \includegraphics[width=\textwidth]{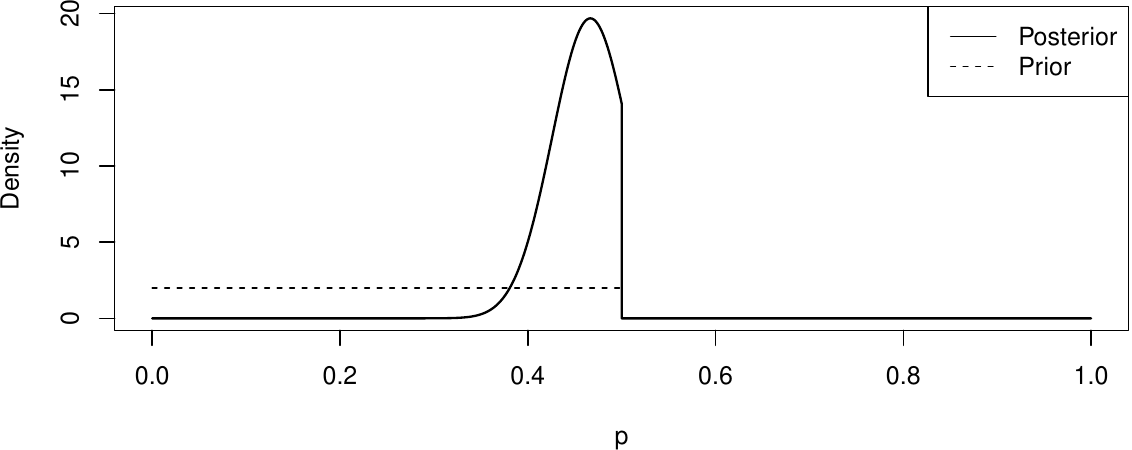}
         \caption{Prior and posterior under $H_0:p\leq \frac{1}{2}$ for the two-sided test of $H_0:p\leq \frac{1}{2}$ against $H_1:p> \frac{1}{2}$ in the therapeutic touch experiment.}
         \label{fig:ex2_1_b}
     \end{subfigure}
     \hfill
     \begin{subfigure}[b]{0.49\textwidth}
         \centering
         \includegraphics[width=\textwidth]{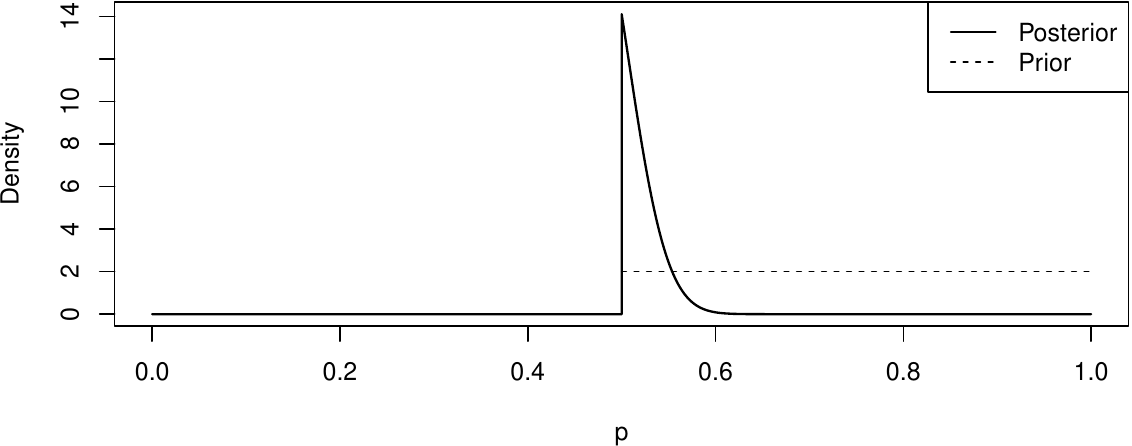}
         \caption{Prior and posterior under $H_1:p > \frac{1}{2}$ for the two-sided test of $H_0:p\leq \frac{1}{2}$ against $H_1:p> \frac{1}{2}$ in the therapeutic touch experiment.}
         \label{fig:ex2_1_c}
     \end{subfigure}
     \hfill
     \begin{subfigure}[b]{0.49\textwidth}
         \centering
         \includegraphics[width=\textwidth]{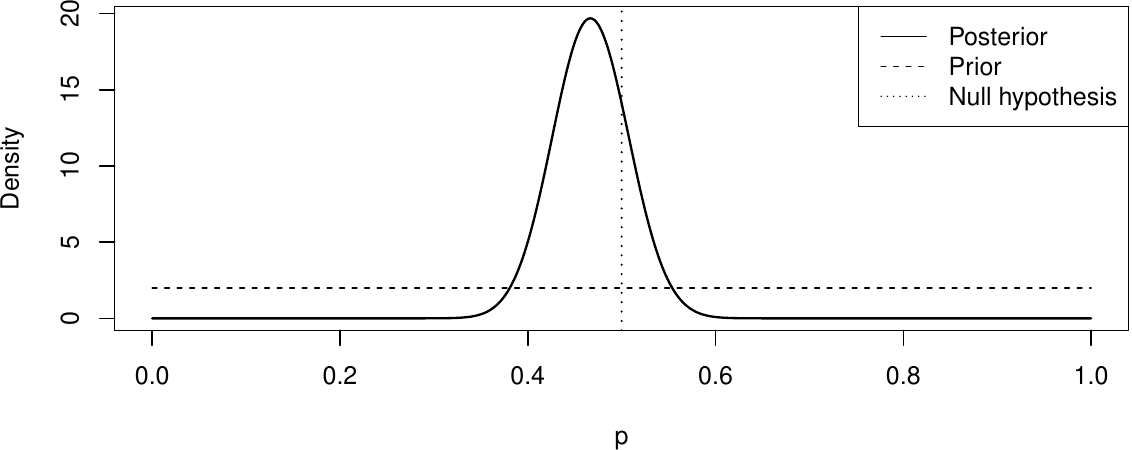}
         \caption{Prior and posterior under $H_0$ and $H_1$ for the two-sided test of $H_0:p\leq \frac{1}{2}$ against $H_1:p> \frac{1}{2}$, combined in one plot.}
         \label{fig:ex2_1_d}
     \end{subfigure}
        \caption{Priors and posteriors for two-sided and directional tests in the therapeutic touch experiment reported by \cite{Rosa1998}.}
    \label{fig:ex_2}
\end{figure}

\Cref{fig:ex_2} shows the associated prior and posterior densities for these two testing scenarios. \Cref{fig:ex2_1_a} shows the prior and posterior for the two-sided test of $H_0:p = \frac{1}{2}$ against $H_1:p \neq \frac{1}{2}$ in the therapeutic touch experiment, where a flat $\mathrm{Beta}(1,1)$ prior is assigned under $H_1$.\footnote{The prior under $H_0$ is again the limiting case of a $\mathrm{Beta}$ prior and reduces to a Dirac measure on $p_0=\frac{1}{2}$, compare \Cref{eq:freqLimit}. The Bayes factor in \Cref{eq:twoSidedBayesFactor} can be read off the plot by making use of the Savage-Dickey density ratio \citep{Verdinelli1995}, and is given as the ratio of the ordinates at the blue and red points.} The Bayes factor in \Cref{eq:twoSidedBayesFactor} results in
\begin{align}\label{eq:ex2_bf_twoSided}
    \mathrm{BF}_{01}(y)=7.05
\end{align}
moderately favouring no effect ($H_0 \colon p = 0.5$) over a therapeutic touch effect ($H_1 \colon p \neq 0.5$).

In comparison, \Cref{fig:ex2_1_b} and \Cref{fig:ex2_1_c} show the resulting prior and posterior under $H_0$ respectively $H_1$, where a truncated $\mathrm{Beta}(1,1)$ prior is used on $H_0$ respectively $H_1$, compare \Cref{eq:analysisPriorsOneSidedTestingH0} and \Cref{eq:analysisPriorsOneSidedTestingH1}. \Cref{fig:ex2_1_d} then shows the plots of \Cref{fig:ex2_1_b} and \Cref{fig:ex2_1_c} blended together, which demonstrates that the priors under $H_0$ and $H_1$ yield a prior on the full parameter space $[0,1]$. The dashed vertical line in \Cref{fig:ex2_1_d} illustrates the boundary $p_0=\frac{1}{2}$ of the null hypothesis $H_0:p\leq \frac{1}{2}$. The resulting Bayes factor for the directional test of the therapeutic touch data results in
\begin{align}\label{eq:ex2_bf_oneSided}
    \mathrm{BF}_{01}(y)=3.81
\end{align}
moderately favouring no or a negative effect ($H_0 \colon p \leq 0.5$) over a positive effect ($H_1 \colon p > 0.5)$, where \Cref{eq:twoSidedBayesFactor} is used to compute $\mathrm{BF}_{01}(y)$.

While computation of a binomial Bayes factor is straightforward, a sample size calculation in advance of the experiment is helpful to gain some insights about the interpretability of the Bayes factor in terms of \Cref{eq:bayesianDesign}. In this setting, we aim for $\alpha=0.05$ and $\beta=0.2$, so we arrive at a Bayesian type-I-error rate of $5\%$ and Bayesian power of at least $80\%$ under the alternative. 

\subsubsection{Directional test of $H_0:p\leq \frac{1}{2}$ against $H_1:p>\frac{1}{2}$}

\begin{figure}[!tb]
    \centering
    
    \begin{subfigure}{0.49\textwidth}
    \includegraphics[width=\linewidth]{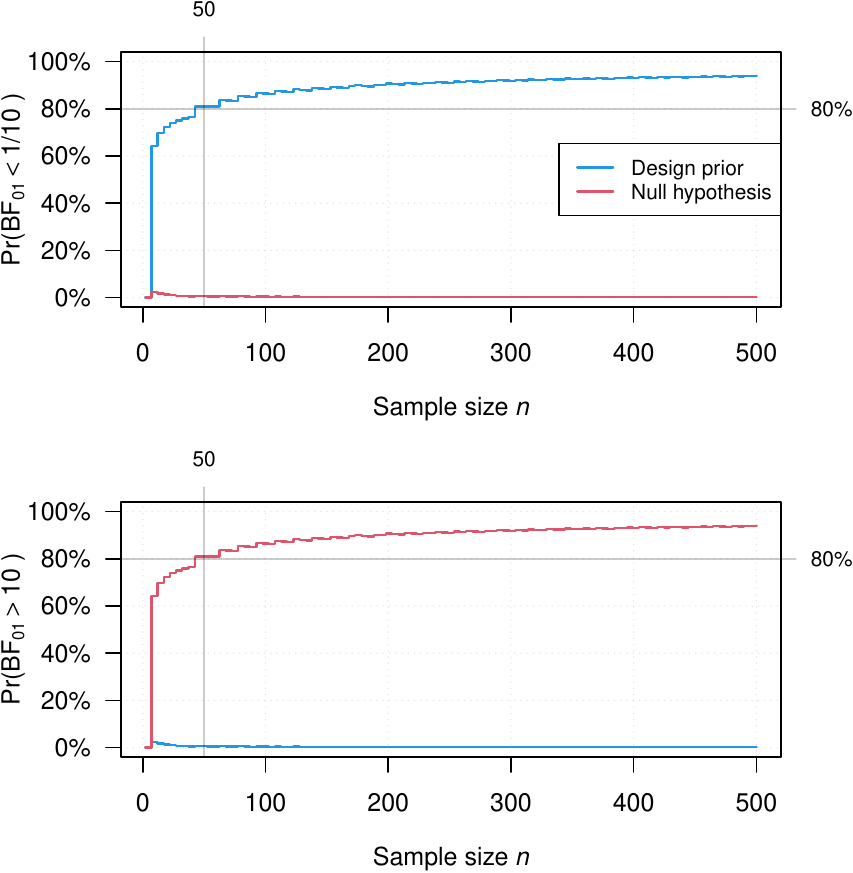}
    \caption{Evidence threshold $k=1/10$}
    \label{fig:ex2_res}
    \end{subfigure}
    \hfill
    \begin{subfigure}{0.49\textwidth}
    \includegraphics[width=\linewidth]{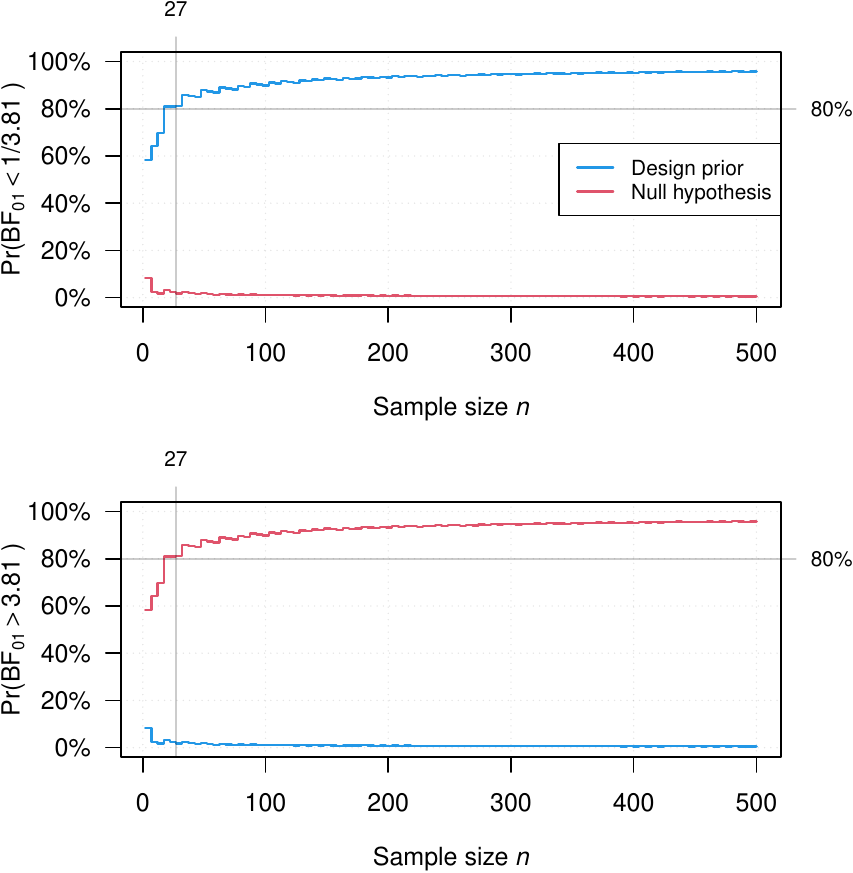}
    \caption{Evidence threshold $k=1/3.81$}
    \label{fig:ex2_res_3.81}
    \end{subfigure}
    
    \caption{Bayesian sample size calculation, power and type-I-error rate (top) and probability of finding compelling evidence for the null hypothesis $H_0$ in the directional test of $H_0:p\leq \frac{1}{2}$ versus $H_1:p>\frac{1}{2}$ of the therapeutic touch experiment reported by \cite{Rosa1998}.}
\end{figure}

First, we perform the directional test, so the alternative hypothesis specifies that the success probability $p$ is larger than $0.5$, thus $H_1:p>0.5$ and $H_0:p\leq 0.5$. The shape of the prior distribution for the Bayes factor -- in our terms the analysis prior -- is specified as the truncated $\mathrm{Beta}(1,1)$ priors under $H_0$ and $H_1$, shown as the dashed horizontal prior densities in \Cref{fig:ex2_1_b} and \Cref{fig:ex2_1_c}. Thus,
$$p \mid H_0 \sim \mathrm{Beta}(1,1)_{[0,0.5]}, \hspace{2cm} p \mid H_1 \sim \mathrm{Beta}(1,1)_{(0.5,1]}$$
We use the same truncated Beta priors as our design priors under $H_0$ and $H_1$, and sample sizes are then calculated via our root-finding approach. The results are shown in \Cref{fig:ex2_res}. 

Based on the flat design priors with $a_d=b_d=1$, $n=50$ participants suffice to attain a power of $81.68\%$ based on the threshold $k=1/10$ of strong evidence. The top plot in \Cref{fig:ex_2} also shows the type-I-error rate under the null hypothesis $H_0:p\leq 0.5$ as the red solid line, and for $n=50$ we arrive at $P(\mathrm{BF}_{01}(y)<1/10 \mid H_0)=0.674\%$. Note that as the design and analysis prior coincide by choice here, the power under the analysis prior used for calculation of the Bayes factors would be identical here.

Another relevant point to mention is what happens if a point prior is used for the design prior, so we attain a frequentist type-I-error rate and power as the limiting case of our Beta prior, compare \Cref{eq:freqLimit}. Then, $P(\mathrm{BF}_{01}(y)<1/10 \mid H_0)$ increases to $P(\mathrm{BF}_{01}(y)<1/10 \mid H_0)=10.13\%$ and the design is not calibrated anymore. 


Taking stock, \Cref{fig:ex2_res} shows that for a convincing Bayesian analysis based on the Bayes factor with evidence threshold $k=1/10$, $n=50$ samples would have sufficed to: 
\begin{enumerate}
    \item[$\blacktriangleright$]{Calibrate the design in terms of the Bayesian type-I-error rate $P(\mathrm{BF}_{01}(y)<k \mid H_0)\leq \alpha$ for $\alpha=0.05$.}
    \item[$\blacktriangleright$]{Calibrate the design in terms of the Bayesian power $P(\mathrm{BF}_{01}(y)<k \mid H_1)> 1-\beta$ for $\beta = 0.8$.}
    \item[$\blacktriangleright$]{Calibrate the design additionally in terms of the Bayesian probability to find compelling evidence for $H_0$ if $H_0$ indeed holds, that is, $P(\mathrm{BF}_{01}(y)>1/k \mid H_0)> 1-\beta$ for $\beta = 0.8$. This is shown in the bottom plot of \Cref{fig:ex2_res}, and holds for the same $n=50$ here.\footnote{This special case is most probably due to the symmetry of $H_0:p\leq 0.5$ and $H_1:p>0.5$ in combination with the flat design and analysis priors.}}
\end{enumerate}

We close the discussion of the directional test in the therapeutic touch experiment by noting that the included number of participants could have been reduced to about $1/3$ of the originally used sample size with our approach, $n=50$ compared to the $n=150$ in the therapeutic touch data.

In addition, this example clarifies another important benefit of our root-finding approach: For trials or experiments already carried out with Bayes factors it becomes possible to calculate power and type-I-error rates retrospectively. This allows to gauge the trustworthiness of a Bayesian test in hindsight based on the reported priors and evidence thresholds, or even to recalculate the entire analysis and check whether sample sizes were large enough to provide reliable conclusions. Here, the directional test resulted in a Bayes factor of $\mathrm{BF}_{01}(y)=3.81$, compare \Cref{eq:ex2_bf_oneSided}. Thus, the Bayes factor does not pass the threshold $k=1/10$. As $n=150$ samples are available in the therapeutic touch experiment and $n=50$ suffice to achieve at least $80\%$ Bayesian power for $H_1:p>\frac{1}{2}$ according to \Cref{fig:ex2_res}, the experimental data fail to provide strong evidence for $H_1$. More importantly, the bottom plot in \Cref{fig:ex2_res} shows that $n=50$ samples also suffice to find compelling evidence for $H_0:p\leq \frac{1}{2}$ if $H_0$ holds. As the resulting Bayes factor is only $\mathrm{BF}_{01}(y)=3.81$, the bottom plot does help little. But we can easily modify the evidence threshold to $k=3.81$, to see if our sample size is large enough to warrant the statement that we have found compelling evidence -- expressed in terms of a Bayes factor as large as $\mathrm{BF}_{01}(y)=3.81$ -- for the null hypothesis, that is, the absence of a therapeutic touch effect.

\Cref{fig:ex2_res_3.81} shows that $n=27$ participants are enough, which is much less than our actually used sample size $n=150$, compare the therapeutic touch data. Thus, if $k=3.81$ is convincing enough in terms of evidential strength, we can argue that a retrospective sample size calculation of the experiment of \cite{Rosa1998} demonstrates the absence of any therapeutic touch effect, that is, compelling evidence for $H_0:p\leq \frac{1}{2}$.\footnote{If the threshold $k=1/3$ for moderate evidence is used, the sample sizes shown in \Cref{fig:ex2_res_3.81} reduce further to $n=22$, much less than the available sample size of $n=150$ in the therapeutic touch data.} Importantly, this statement is backed by the power and type-I-error guarantees expressed in \Cref{eq:bayesianDesign}, that is, if there had been a therapeutic touch effect, we had at least $80\%$ power to detect it. Also, the probability of a false-positive result was bounded by $5\%$.

\subsubsection{Two-sided test of $H_0:p = \frac{1}{2}$ against $H_1:p \neq \frac{1}{2}$}

\begin{figure}[h!]
    \centering
    
    \begin{subfigure}{0.495\textwidth}
    \includegraphics[width=1\linewidth]{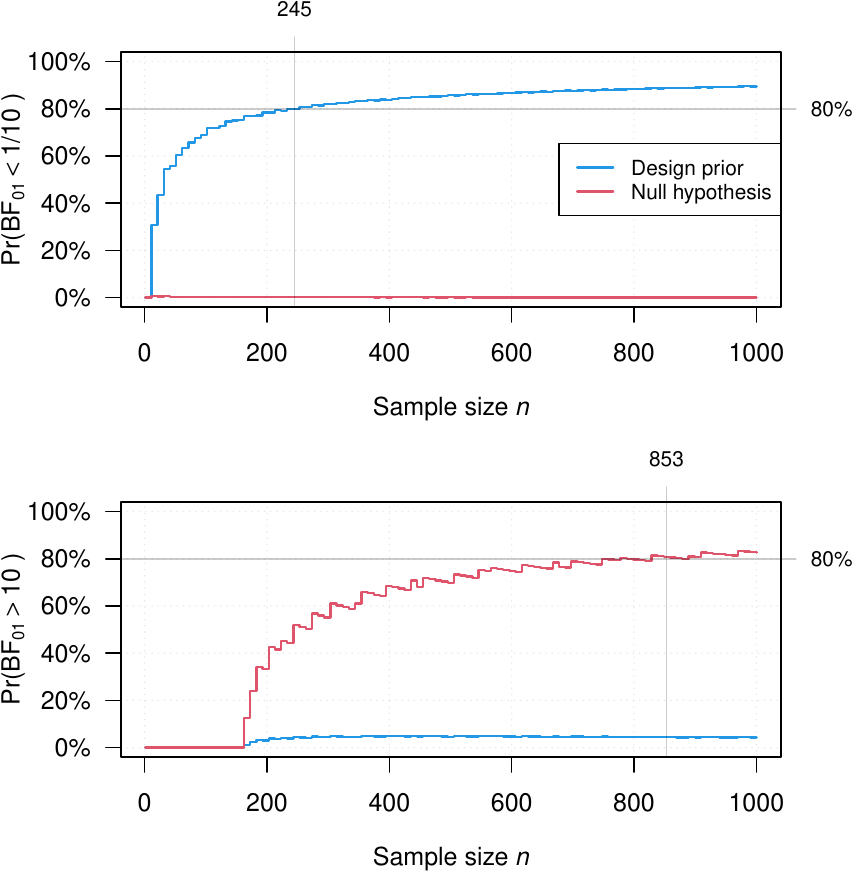}
    \caption{Evidence threshold $k=1/10$}
    \label{fig:ex2_res_twoSided}
    \end{subfigure}
    \hfill
    \begin{subfigure}{0.495\textwidth}
    \includegraphics[width=1\linewidth]{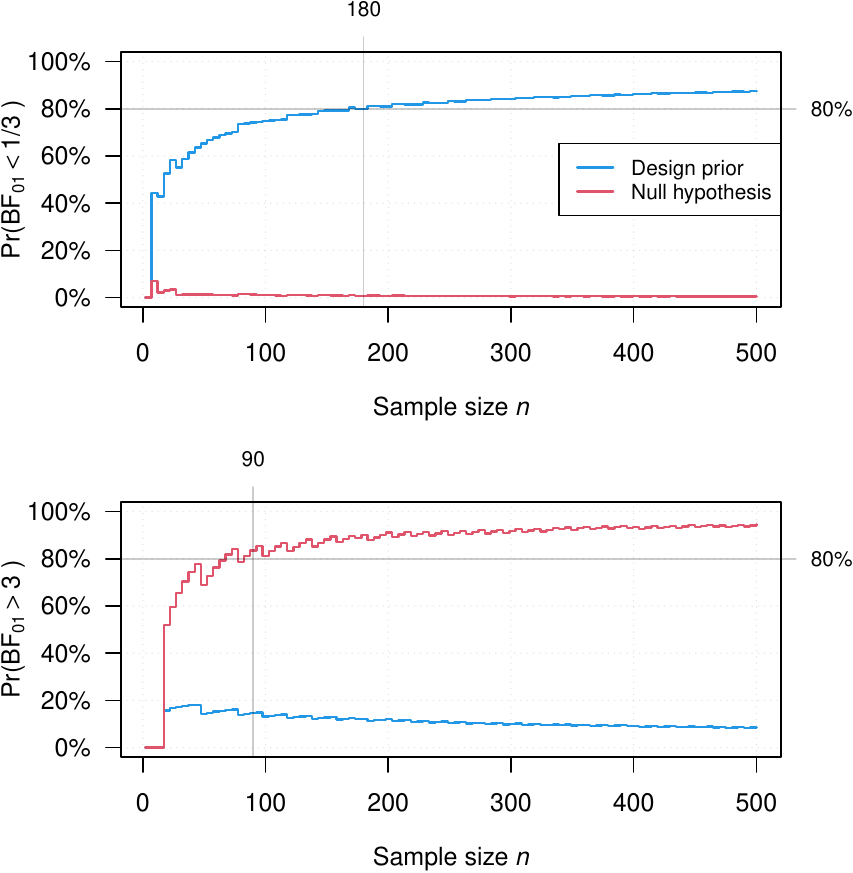}
    \caption{Evidence threshold $k=1/3$}
    \label{fig:ex2_res_twoSided_moderate}
    \end{subfigure}
    
    \caption{Bayesian sample size calculation, power and type-I-error rate (top) and probability of finding compelling evidence for the null hypothesis $H_0$ in the two-sided test of $H_0:p= \frac{1}{2}$ versus $H_1:p \neq \frac{1}{2}$ of the therapeutic touch experiment reported by \cite{Rosa1998}.}
\end{figure}

We turn to the two-sided test in the experiment of \cite{Rosa1998} now. Thus, we test $H_0:p=\frac{1}{2}$ versus $H_1:p\neq \frac{1}{2}$ and use the prior shown in \Cref{fig:ex2_1_a}. Under $H_1$, we employ a flat Beta $\mathrm{Beta}(1,1)$ prior, while under $H_0$, we put a Dirac measure on $p=\frac{1}{2}$. The design and analysis priors under $H_1$ are chosen identically as the flat Beta prior, and \Cref{fig:ex2_res_twoSided} shows the results of the root-finding approach to Bayesian sample size calculation.

Now the required number of participants increases to $n=245$ to attain a Bayesian power of at least $80\%$. The reason is that now $H_0$ and $H_1$ are less well separated as parameter values slightly larger or smaller than $p_0=\frac{1}{2}$ are included in the design prior under $H_1$. This leads to a larger sample size than in the directional test. 


The required sample size to find compelling evidence for $H_0:p=\frac{1}{2}$ shown in the bottom plot of \Cref{fig:ex2_res_twoSided} increases to $n=853$ samples. Reducing the evidence threshold to $k=1/3$ yields the results shown in \Cref{fig:ex2_res_twoSided_moderate}, and now $n=180$ suffices to attain a Bayesian power of at least $80\%$ under $H_1:p\neq \frac{1}{2}$, whereas the probability to find compelling evidence for $H_0:p=\frac{1}{2}$ passes the threshold of $80\%$ at $n=90$ participants, compare the bottom plot in \Cref{fig:ex2_res_twoSided_moderate}.

Note that the Bayes factor for the two-sided test resulted in $\mathrm{BF}_{01}(y)=7.05$, compare \Cref{eq:ex2_bf_twoSided}. Thus, based both $k=1/10$ or $k=1/3$, the top plots of \Cref{fig:ex2_res_twoSided} and \Cref{fig:ex2_res_twoSided_moderate} clarify that $n=150$ samples are not enough to achieve $80\%$ Bayesian power for $H_1:p\neq \frac{1}{2}$, that is, in favour of a therapeutic touch effect. Also, the bottom plot of \Cref{fig:ex2_res_twoSided} shows that the sample size of $n=150$ is way too small to be able to demonstrate compelling evidence in favour of $H_0:p=\frac{1}{2}$ based on the strong evidence threshold $k=1/10$. However, the bottom plot in \Cref{fig:ex2_res_twoSided_moderate} shows that if the evidence threshold is reduced to $k=1/3$, $n=90$ samples suffice to achieve $80\%$ probability to find compelling evidence for the null hypothesis $H_0:p=\frac{1}{2}$, if $H_0$ indeed holds. As $n=150>90$, see the therapeutic touch data, and $\mathrm{BF}_{01}(y)=7.05>3$, the Bayesian sample size calculation in \Cref{fig:ex2_res_twoSided_moderate} demonstrates that the experimental sample size was large enough to find a Bayes factor of at least $3$ for $H_0$ with at least $80\%$ probability. From this perspective, a retrospective Bayesian power analysis with our root-finding approach demonstrates that the experiment was designed well enough to interpret the Bayes factor $\mathrm{BF}_{01}(y)=7.05>3$ as moderate evidence in favour of $H_0:p=\frac{1}{2}$. However, as $n=180$ samples would have been necessary for $80\%$ Bayesian power under $H_1$ and only $n=150$ were available, the study was underpowered for $H_1:p>\frac{1}{2}$ when the two-sided test is used.

We can even calculate how large the Bayesian power for e.g. $k=1/3$ or $k=1/10$ would have been to achieve a Bayes factor $\mathrm{BF}_{01}<1/3$ or $\mathrm{BF}_{01}<1/10$ based on our $n=150$ samples of the therapeutic touch data: The root-finding approach yields $P(\mathrm{BF}_{01}(y)<1/10) \mid H_1,n=150)=75.50\%$, whereas $P(\mathrm{BF}_{01}(y)<1/10) \mid H_1,n=150)=79.47\%$. Thus, the Bayesian power to obtain a Bayes factor $<1/10$ respectively $<1/3$ was quite large in the experiment. Despite this fact, the experimental data failed to achieve a Bayes factor $\mathrm{BF}_{01}$ anywhere close to $1/10$ or $1/3$. Thus, our root-finding approach clarifies that even in the underpowered setting, the power to find a large enough Bayes factor was quite close to the desired $80\%$, backing the statement that there is no therapeutic touch effect.\footnote{We stress that these power calculations are based on a flat design prior. If the true effect is very small, perhaps $p=0.51$, the study would have been very much underpowered. However, in the state of equipoise about the truth of $H_0$ and $H_1$, the flat design prior with $a_d=b_d=1$ is a reasonable choice, and the power calculation based on such a prior seems justified.}


\section{Magic mirror on the wall, which is the best metric of them all?}\label{sec:powerConcepts}
A key question in the design of experiments is which metric to use to calibrate a design for some statistical test. In \Cref{eq:bayesianDesign}, we formulated Bayesian analogues to the Neyman-Pearsonian school of thought, where the type-I-error rate is bounded by some $\alpha \in (0,1)$ and the power is maximized subsequently under $H_1$. Here, we formulated a bound on the Bayesian type-I-error rate as $P(\mathrm{BF}_{01}(y)<k|H_0)<\alpha$, and the requirement of at least $1-\beta$ Bayesian power under $H_1$ as $P(\mathrm{BF}_{01}(y)<k|H_1)>1-\beta$ for some $\beta \in (0,1)$. Using this metric comes close to a Bayes-frequentist compromise in the spirit of \cite{Good1983a}, who argued that the distribution of Bayes factors (under $H_0$ and $H_1$) can be used by a frequentist as a test statistic. However, a key advantage of our approach as shown in the examples in \Cref{sec:examples} is that we also can calibrate the probability to find compelling evidence for the null hypothesis $H_0$. That is, we can design our study to fulfill the bound
\begin{align}
    P(\mathrm{BF}_{01}(y)>1/k|H_0)>1-\beta'
\end{align}
for some $\beta'\in (0,1)$. For $k=1/10$, the latter amounts to design the experiment so that we have a probability of at least $1-\beta'$ to find at least strong evidence in favour of $H_0$, based on the Bayes factor. As discussed in the phase II clinical trial example, in a variety of situations it might be highly desirable to design a study to have a large probability to find compelling evidence for $H_0$, if $H_0$ indeed holds. In the bottom part of \Cref{fig:overview}, the three core metrics are shown which are already implemented in the \texttt{bfpwr} package.

In summary, these are:
\begin{enumerate}
    \item[$\blacktriangleright$]{The \textit{Bayesian type-I-error rate} $P(\mathrm{BF}_{01}<k|H_0)$, where the smallest $n\in \mathbb{N}$ is calculated by root-finding for which $P(\mathrm{BF}_{01}<k|H_0)\leq\alpha$ for some $\alpha \in (0,1)$, e.g. $\alpha=0.05$ or $\alpha=0.025$.}
    \item[$\blacktriangleright$]{The \textit{Bayesian power} $P(\mathrm{BF}_{01}<k|H_1)$, where the smallest $n\in \mathbb{N}$ is calculated by root-finding for which $P(\mathrm{BF}_{01}<k|H_0)>1-\beta$ for some $\beta \in (0,1)$, e.g. $\beta=0.2$ or $\beta=0.1$.}
    \item[$\blacktriangleright$]{The Bayesian probability of compelling evidence for $H_0$, $P(\mathrm{BF}_{01}>1/k|H_0)$, where the smallest $n\in \mathbb{N}$ is calculated by root-finding for which $P(\mathrm{BF}_{01}>1/k|H_0)>\beta'$ for some $\beta' \in (0,1)$, e.g. $\beta'=0.2$ or $\beta'=0.1$.}
\end{enumerate}

We stress that calibrating according to a metric is like a modular system with our root-finding approach: We can calibrate an experimental design according to the Bayesian type-I-error rate, the Bayesian power, the probability to find compelling evidence for the null hypothesis, or combinations thereof. For example, there might be situations where we are only interested in Bayesian power under $H_1$, and the root-finding approach yields $n_1$ required samples. In another situation, a type-I-error guarantee is required (e.g. due to the requirement for an ethics committee or a trial steering commitee), and the root-finding approach yields $n_2$ required samples now. Taking $\max(n_1,n_2)$ then yields a design which is calibrated for both aspects.

There might even be different metrics for the Bayes factor -- or another test statistic, see the discussion section about possible generalizations of our root-finding approach -- that could be of interest and are required to be calibrated. For example, we could interpret indecisive Bayes factors between $1/3$ and $3$ as undesirable and formulate the metric
$$P(1/3<\mathrm{BF}_{01}(y)<3|H_i)<\alpha'$$
for some $\alpha'\in (0,1)$, which could be termed as bounding the probability of indecisive evidence under $H_i$, $i=0,1$.
Extending our three core metrics to such scenarios and carrying out the root-finding approach is still possible then, although not implemented yet.

We close this section by noting that the answer to the question 'Magic mirror on the wall, which is the best metric of them all?' depends on the requirements of researchers at the end, our root-finding approach allows for a flexible calibration of a Bayesian design for a variety of metrics which could possibly be of interest in applied research.



\section{Discussion}\label{sec:discussion}

Bayesian design of experiments and sample size calculations usually rely on complex Monte Carlo simulations in practice. Obtaining bounds on Bayesian notions of the false-positive rate and power therefore often lack closed-form or approximate numerical solutions. In this paper, we introduced a novel approach to Bayesian sample size calculation in the binomial setting via Bayes factors. We discussed the drawbacks of sample size calculations via Monte Carlo simulations and proposed a numerical root-finding approach which allows to determine the necessary sample size to obtain prespecified bounds of Bayesian power and type-I-error rate almost instantaneously. Real-world examples and applications in clinical trials illustrated the advantage of the proposed method, where we focussed on point-null versus composite and directional hypothesis tests. We derived the corresponding Bayes factors and discussed relevant aspects to consider when pursuing Bayesian design of experiments with the introduced approach. In summary, our approach allows for a Bayes-frequentist compromise by providing a Bayesian analogue to a frequentist power analysis for the Bayes factor in binomial settings. The methods are implemented in our R package \texttt{bfpwr}\footnote{The binomial test methods are currently only on the development version which can be installed by running \texttt{remotes::install\_github(repo = "SamCH93/bfpwr", subdir = "package", ref = "binomial")}, which requires the \texttt{remotes} package. They will soon be made available in the CRAN version of the package, which can be installed with \texttt{install.packages("bfpwr")}.}.

In this section, we discuss several points which were only addressed briefly and elaborate on extensions and future work in this direction.

\subsection{Generalization to other hypothesis testing approaches}
An important point not discussed in detail so far is inherent in the workflow depicted in \Cref{fig:overview}: Our root-finding approach is not limited to sample size calculations for Bayes factors. In fact, posterior probabilities could also be used by replacing \Cref{eq:rootfinding} with
\begin{align}
    P_{p \mid Y}(H_1)=k
\end{align}
where now $k$ has a different interpretation and $k\in (0,1)$.The solution can be found by simple numerical methods, and the above clarifies that the root-finding approach is extendable to any statistical test as long as a numerical root-finding method is available. This could become difficult in larger dimensions, but for most statistical tests there exist well-established numerical methods to solve the root in moderately-dimensional parameter spaces.

Other examples next to posterior probability of a hypothesis are the posterior odds $\frac{P_{p \mid Y}(H_1)}{P_{p \mid Y}(H_0)}$, or other Bayesian indices for hypothesis testing like the region of practical equivalence (ROPE) or Bayesian evidence values, see for example \cite{Kelter2020BayesianPosteriorIndices,Kelter2022EvidenceValue}.

\subsection{Retrospective or reverse-Bayes sample size calculations for Bayes factors}

The therapeutic touch experiment discussed in detail in \Cref{sec:examples} clarified another important benefit of our root-finding approach: For trials or experiments already carried out with Bayes factors it becomes possible to calculate power and type-I-error rates retrospectively with our approach. This allows to gauge the trustworthiness of a Bayesian test in hindsight based on the reported priors and evidence thresholds, or even to recalculate the entire analysis and check whether sample sizes were large enough to provide reliable conclusions. A reverse-Bayes analysis in the spirit of \cite{Good1983a} is also possible, which answers the question:
\begin{quote}
    Which design prior do we need to achieve a calibrated design?
\end{quote}
See also \citet{Held2021} for a review of reverse-Bayes approaches. Another type of retrospective analysis is possible with our method, too, and also based on the detailed discussion of the therapeutic touch example:
\begin{quote}
    How small must the evidence threshold $k$ be so that one can accept the resulting Bayes factor in the experiment carried out with sample size $n$ as a convincing analysis in the sense that it is based on a \textit{calibrated} design according to \Cref{eq:bayesianDesign}?
\end{quote}
If one is willing to accept this threshold $k$, an analysis is trustworthy in hindsight. If it is not, the reverse-Bayes analysis casts doubt on the trustworthiness of the carried out experiment. Thus, this second type of retrospective analysis allows to find the evidence threshold $k$ for which based on the experimental sample size $n$ the design would have been calibrated. Based on the reported Bayes factor $\mathrm{BF}_{01}(y)$ one can then check whether $\mathrm{BF}_{01}(y)<k$, and if so, the analysis is calibrated and trustworthy in hindsight.

\subsection{Generalization to other settings}

Most importantly, the root-finding approach is not limited to the binomial setting covered in detail in this paper. It is possible to use the same approach for other tests, including $z$ tests or $t$ tests, compare \cite{PawelHeld2024} and \cite{WongTendeiro2024}. As long as the Bayes factor can be found numerically and a numerical or even analytic solution to \Cref{eq:rootfinding} is possible, root-finding of the necessary sample size to achieve a prespecified Bayesian power and type-I-error rate should, in general, be possible. The advantage of our approach is obvious in this regard: The sample size calculations require almost no computation time and work almost instantaneously. This is in sharp contrast to the current standard of performing a Monte Carlo simulation study to reveal the required sample size in a given test problem, whether posterior probabilities, Bayes factors or posterior odds are the chosen test criterion.

\appendix
\appendixpage

\section*{Proofs}\label{sec:proofs}

\begin{proof}{Expectation of a truncated beta random variable}

    Let $X$ be a beta random variable truncated to the interval from $l$ to $u$, i.e., $X \sim \mathrm{Beta}(a,b)_{[l,u]}$. The expectation of $X$ is
    
    \begin{align}
        \mathrm{E}[X] 
        &= \int_l^u x \, \frac{x^{a-1} \, (1 - x)^{b - 1}}{\mathrm{B}(a, b) \, \{I_u(a, b) - I_l(a, b)\}} \, \mathrm{d}x \nonumber \\
        &= \int_l^u \frac{x^{a + 1 -1} \, (1 - x)^{b - 1}}{\mathrm{B}(a, b) \, \{I_u(a, b) - I_l(a, b)\}} \, \mathrm{d}x \nonumber \\
        &= \frac{\mathrm{B}(a + 1, b)}{\mathrm{B}(a, b) \, \{I_u(a, b) - I_l(a, b)\}} \underbrace{\int_l^u \frac{x^{a + 1 -1} \, (1 - x)^{b - 1}}{\mathrm{B}(a + 1, b)} \, \mathrm{d}x}_{=I_u(a+1, b) - I_l(a+1, b)} \nonumber\\
        &= \frac{\mathrm{B}(a + 1, b)}{\mathrm{B}(a, b)} \times \frac{I_u(a+1, b) - I_l(a+1, b)}{I_u(a, b) - I_l(a, b)} \nonumber\\
        &= \frac{a}{a + b} \times \frac{I_u(a+1, b) - I_l(a+1, b)}{I_u(a, b) - I_l(a, b)} \label{eq:expectationTruncatedBeta}
    \end{align}
    where the last equality follows from Pascal's identity. 
\end{proof}
\medskip

\begin{proof}[Derivation of the prior-predictive probability mass function]
First, the prior-predictive is given by definition as
\begin{align}
    f(y \mid n, a, b, l, u) 
        &= \frac{\displaystyle \int_l^u \mathrm{Bin}(y \mid n, p) \, \mathrm{Beta}(p \mid a, b) \, \mathrm{d} p}{I_{u}(a, b) - I_{l}(a, b)}\nonumber\\
        &=\frac{\displaystyle \int_l^u {n\choose y} p^y (1-p)^{n-y} p^{a-1}(1-p)^{b-1}dp}{\mathrm{B}(a,b)[I_u(a,b)-I_l(a,b)]}\nonumber\\
        &={n\choose y}\frac{1}{\mathrm{B}(a,b)[I_u(a,b)-I_l(a,b)]}\displaystyle \int_l^u  p^{y+a-1} (1-p)^{n+b-y-1} dp \label{eq:priorPredDeriv1}
\end{align}
and from
\begin{align}\label{eq:priorPredDeriv2}
    \int_l^u  p^{y+a-1} (1-p)^{n+b-y-1} dp=\mathrm{B}(a+b,b+n-y)\cdot [I_u(a+y,b+n-y)-I_l(a+y,b+n-y)]
\end{align}
it follows by substituting \Cref{eq:priorPredDeriv2} in \Cref{eq:priorPredDeriv1} that
\begin{align}
    f(y \mid n, a, b, l, u) ={n\choose y}\frac{\mathrm{B}(a+b,b+n-y)[I_u(a+y,b+n-y)-I_l(a+y,b+n-y)]}{\mathrm{B}(a,b)[I_u(a,b)-I_l(a,b)]}
\end{align}
which is \Cref{eq:predDensity}.
\end{proof}

\begin{proof}[Derivation of the Bayes factor for the one-sided test of $H_0:p\leq p_0$ versus $H_1:p>p_0$ for $p_0\in (0,1)$]
By definition, the Bayes factor is given as
\begin{align}
    \mathrm{BF}_{01}(y)=\frac{f(y|H_0)}{f(y|H_0)}
\end{align}
Based on the truncated Beta analysis prior, compare \Cref{eq:analysisPriorsOneSidedTestingH0} and \Cref{eq:analysisPriorsOneSidedTestingH1}, the predictive probability mass functions under $H_0$ and $H_1$ are given as
\begin{align}
    f(y|H_0)=\int_{0}^{p_0}\mathrm{Bin}(y|n,p)\mathrm{Beta}_{[0,0.2]}(p|a_a,b_a)dp
\end{align}
and 
\begin{align}
    f(y|H_1)=\int_{p_0}^{1}\mathrm{Bin}(y|n,p)\mathrm{Beta}_{[0.2,1]}(p|a_a,b_a)dp
\end{align}
so 
\begin{align}
    \mathrm{BF}_{01}(y)=\frac{f(y|H_0)}{f(y|H_0)}=&\frac{\displaystyle\int_{0}^{p_0}\mathrm{Bin}(y|n,p)\mathrm{Beta}_{[0,0.2]}(p|a_a,b_a)dp}{\displaystyle\int_{p_0}^{1}\mathrm{Bin}(y|n,p)\mathrm{Beta}_{[0.2,1]}(p|a_a,b_a)dp}\\
    &=\frac{{n\choose y}}{{n\choose y}}\cdot \frac{\displaystyle \int_{0}^{p_0}p^y (1-p)^{n-y}\frac{p^{a_a-1}(1-p)^{b_a-1}}{\mathrm{B}(a_a,b_a)I_{p_0}(a_a,b_a)}dp}{\displaystyle \int_{p_0}^{1}p^y (1-p)^{n-y}\frac{p^{a_a-1}(1-p)^{b_b-1}}{\mathrm{B}(a_a,b_a)[I_{1}(a_a,b_a)-I_{p_0}(a_a,b_a)]}dp}\label{eq:proofBFOneSided}\\
\end{align}
From
$$I_1(a_a,b_a)=\frac{\mathrm{B}(1;a_a,b_a)}{\mathrm{B}(a_a,b_a)}=1$$
the right-hand side of Equation \Cref{eq:proofBFOneSided} simplifies further to
\begin{align}
    &\frac{\displaystyle \int_{0}^{p_0}p^y (1-p)^{n-y}\frac{p^{a_a-1}(1-p)^{b_a-1}}{\mathrm{B}(a_a,b_a)I_{p_0}(a_a,b_a)}dp}{\displaystyle \int_{p_0}^{1}p^y (1-p)^{n-y}\frac{p^{a_a-1}(1-p)^{b_b-1}}{\mathrm{B}(a_a,b_a)[1-I_{p_0}(a_a,b_a)]}dp}\label{eq:proofBFOneSided2}\\
    &=\frac{\displaystyle \int_{0}^{p_0}p^{y+a_a-1}(1-p)^{n-y+b_a-1}dp\cdot \frac{1}{\mathrm{B}(a_a,b_a)}}{\displaystyle \int_{p_0}^{1}p^{y+a_a-1}(1-p)^{n-y+b_a-1}dp\cdot \frac{1}{\mathrm{B}(a_a,b_a)}}\cdot \frac{1-I_{p_0}(a_a,b_a)}{I_{p_0}(a_a,b_a)}\label{eq:proofBFOneSided3}
\end{align}
The numerator of \Cref{eq:proofBFOneSided3} can be written as
\begin{align}
    \displaystyle \int_{0}^{p_0}p^{y+a_a-1}(1-p)^{n-y+b_a-1}dp\cdot \frac{1}{\mathrm{B}(a_a,b_a)}&=I_{p_0}(a_a+y,b_a+n-y)-\underbrace{I_{0}(a_a+y,b_a+n-y)}_{=0}\nonumber\\
&=I_{p_0}(a_a+y,b_a+n-y)
\end{align}
and the denominator in \Cref{eq:proofBFOneSided3} can be written as
\begin{align}
    \displaystyle \int_{p_0}^{1}p^{y+a_a-1}(1-p)^{n-y+b_a-1}dp\cdot \frac{1}{\mathrm{B}(a_a,b_a)}&=I_{1}(a_a+y,b_a+n-y)-I_{p_0}(a_a+y,b_a+n-y)\nonumber\\
    &=1-I_{p_0}(a_a+y,b_a+n-y)
\end{align}
and therefore \Cref{eq:proofBFOneSided3} simplifies further to
\begin{align}
    &\frac{\displaystyle \int_{0}^{p_0}p^{y+a_a-1}(1-p)^{n-y+b_a-1}dp\cdot \frac{1}{\mathrm{B}(a_a,b_a)}}{\displaystyle \int_{p_0}^{1}p^{y+a_a-1}(1-p)^{n-y+b_a-1}dp\cdot \frac{1}{\mathrm{B}(a_a,b_a)}}\cdot \frac{1-I_{p_0}(a_a,b_a)}{I_{p_0}(a_a,b_a)}\nonumber\\
    &=\frac{I_{p_0}(a_a+y,b_a+n-y)\cdot (1-I_{p_0}(a_a,b_a))}{1-I_{p_0}(a_a+y,b_a+n-y)\cdot (I_{p_0}(a_a,b_a))}\label{eq:proofBFOneSided4}
\end{align}
and the right-hand side of \Cref{eq:proofBFOneSided4} yields \Cref{eq:bayesFactorOneSided} and finishes the proof.
\end{proof}

\begin{proof}[Derivation of the Bayes factor for the two-sided test of $H_0:p=p_0$ versus $H_1:p\neq p_0$]
Based on the same assumption $Y \mid p \sim \mathrm{Bin}(n, p)$, a Dirac measure in $p_0$ is chosen as the analysis prior under $H_0$, that is,
$$p \mid H_0 \sim \delta_{p_0}$$. A Beta analysis prior is chosen under $H_1$, that is, $$p \mid H_1 \sim \mathrm{Beta}(a_a,b_a)$$
with hyperparameters $a_a,b_a$. Based on these analysis priors under $H_0$ and $H_1$, we have
\begin{align}
        \mathrm{BF}_{01}(y) = \frac{f(y \mid H_0)}{f(y \mid H_1)} &= \frac{{n\choose y}p_0^y \, (1 - p_0)^{n - y}}{\displaystyle\int_{0}^{1}{n\choose y}p^y \, (1 - p)^{n - y} \frac{p^{a-1}(1-p)^{b-1}}{B(a,b)}dp}\\
        &=\frac{\bcancel{{n\choose y}}p_0^y \, (1 - p_0)^{n - y}}{\displaystyle\int_{0}^{1}\bcancel{{n\choose y}}p^y \, (1 - p)^{n - y} \frac{p^{a-1}(1-p)^{b-1}}{B(a,b)}dp}\\
        &=\frac{p_0^y \, (1 - p_0)^{n - y} B(a,b)}{\displaystyle\int_{0}^{1}p^{y+a-1} \, (1 - p)^{b + n - y -1}dp}\\
        & = p_0^y \, (1 - p_0)^{n - y} \, \frac{\mathrm{B}(a , b)}{\mathrm{B}(a + y, b + n - y)}\label{eq:proofBFTwoSided}
    \end{align}
    where the right-hand side of \Cref{eq:proofBFTwoSided} yields \Cref{eq:twoSidedBayesFactor}.
\end{proof}

\begin{proof}[Relationship of $a_d$ and $b_d$ for mode-centered Beta design priors]

Let $X$ be a beta random variable, then the mode is given as $\frac{a_d-1}{a_d+b_d-2}$ for $a_d,b_d>1$, or as any value in the domain of $X$ for $a_d=b_d=1$. Let $p_1>p_0$ be given for some $p_1$ in $H_1$, then 
\begin{align}
    &\frac{a_d-1}{a_d+b_d-2} = p_1 \nonumber\\
    &\Leftrightarrow a_d-1 = p_0[a_d+b_d-2] \nonumber \\
     &\Leftrightarrow(1-p_1)a_d-1=p_1 b_d -2p_1 \label{eq:proofMeanModeCentering}\\
     &\Leftrightarrow a_d = \frac{p_1(b_d - 2) + 1}{1 - p_1} \nonumber
\end{align}
So for any $b_d>1$, selection of $a_d$ as specified above yields a mode-centered Beta random variable. As the mode is invariant under truncation, because each density value is normalized only with a multiplicative constant, this also holds for the truncated Beta design priors.
\end{proof}

\begin{proof}[Mean- and mode-centering for a Beta random variable]
Now, the mean of a Beta random variable $X\sim \mathrm{Beta}(a_d,b_d)$ is given as $\frac{a_d}{a_d+b_d}$, and from
\begin{align}
    \frac{a_d}{a_d+b_d}=p_1
\end{align}
for some $p_1>p_0$ from $H_1$, we obtain
\begin{align}
    a_d = p_1(a_d+b_d) \Leftrightarrow (1-p_1)a_d = p_1 b_d \Leftrightarrow b_d = \frac{(1-p_1)a_d}{p_1}
\end{align}
Substitution of the latter in \Cref{eq:proofMeanModeCentering} yields
\begin{align}
    (1-p_1)a_d-1=(1-p_1)a_d-2p_1 \Leftrightarrow -1 = -2p_1 \Leftrightarrow p_1 = \frac{1}{2}
\end{align}
which proves that mean- and mode-centering is possible only when $p=\frac{1}{2}$. Note that while the mean changes when truncating the Beta design priors, the mode stays invariant, so the proof also works when using the expectation of the truncated Beta distribution, because the multiplicative constant in the right-hand side of \Cref{eq:expectationTruncatedBeta} only modifies the resulting value for $p_1$.
\end{proof}

\bibliography{library.bib} 

\end{document}